\documentclass[aps,pra,twocolumn,showpacs,superscriptaddress,floatfix]{revtex4-1}
\usepackage{array}
\usepackage{graphicx,amsmath,amssymb}
\usepackage[usenames]{color}
\usepackage[dvipsnames]{xcolor}
\usepackage[colorlinks=true,linkcolor=blue,anchorcolor=blue,citecolor=blue,urlcolor=blue]{hyperref}
\begin{document}


\title{Enhanced two-photon blockade without cascade channel by Stark nonlinear coupling}

\author{Zu-Jian Ying}
\email{yingzj@lzu.edu.cn} 
\affiliation{School of Physical Science and Technology, Lanzhou University, Lanzhou 730000, China}
\affiliation{Key Laboratory for Quantum Theory and Applications of MoE, Lanzhou Center for Theoretical Physics, Lanzhou University, Lanzhou 730000, China}

\author{Hang-Hang Han}
\affiliation{School of Physical Science and Technology, Lanzhou University, Lanzhou 730000, China}
\affiliation{Key Laboratory for Quantum Theory and Applications of MoE, Lanzhou Center for Theoretical Physics, Lanzhou University, Lanzhou 730000, China}

\begin{abstract}
Two-photon blockade (TPB) besides the conventional one-photon blockade can control the photon at the level of individual quanta in the input-output measurements of light-matter coupling systems. Apart from conventional TPB (C-TPB) with opened cascade decay channel, unconventional TPB (U-TPB) with closed cascade decay channel may open novel avenues for manipulation of TPB. However, currently found U-TPB in linear coupling is limited in a narrow coupling window and the blockade strength is weak, which would hinder its applications. In the present work we propose to enhance the U-TPB by the Stark nonlinear coupling. Indeed, the introduction of the Stark nonlinear coupling to the linear coupling enables tuning of both cascade energy level and the anharmonicity which play key roles in the formation of TPBs.  By investigating photon correlation functions and extracting phase diagrams in dissipation, we demonstrate that our scheme not only dramatically broadens the window of U-TPB to cover the entire strong-coupling regime but also deepens the blockading degree of the U-TPB by two orders. The cascade decay rates, state populations and the anharmonicity, are examined to identify and track the U-TPB and C-TPB. In the overview of phase diagrams we also reveal a broad U-TPB in ultrastrong couplings, with a crossover to the C-TPB. Since the Stark nonlinear coupling is realizable and tailorable, our proposal may pave a practical way for manipulation of the TPB.
\end{abstract}

\maketitle

\section{Introduction}

Developments of advanced quantum technologies have become one of the central
topics in modern physics. In such a contemporary trend, quantum metrology~\cite{Garbe2020,Montenegro2021-Metrology,Chu2021-Metrology,Garbe2021-Metrology,Ilias2022-Metrology, Ying2022-Metrology,Gietka2023PRL-Squeezing,YangZheng2023SciChina,Hotter2024-Metrology,Alushi2024PRL,Mukhopadhyay2024PRL,Mihailescuy2024, Ying-Topo-JC-nonHermitian-Fisher,*Ying-Topo-JC-nonHermitian-Fisher-Cover,Ying-g2hz-QFI-2024,*Ying-g2hz-QFI-2024-Cover, Ying-g1g2hz-QFI-2025,Ying2025g2A4,*Ying2025g2A4-Cover,Ying-g2Stark-QFI-2025,Gietka2025PRL100802,Mihailescu2025CQMtutorial,QiaoFeng2026SpinSqueeze,QiuYi2025gA2}
and quantum manipulation~\cite{Boite2016-Photon-Blockade,Ridolfo2012-Photon-Blockade,LiaoJQ2020BlockadeJC,Ma2026PRL-Strong-2PB,Felicetti2026PRXQuantumBlockade, Carmichael1985,Birnbaum2005,Shamailov2010,Liew2010,Hamsen2017,Garziano2017,Flayac2017,Han2026WeakTPB,
Nagasawa2013Rings,Ying2016Ellipse,Ying2017curvedSC,Ying2020PRR,Gentile2022NatElec} are two crucial directions. Essentially, the former
pursues ultrahigh measurement precision by exploiting quantum resources
and the latter aims to control the quantum resources, in all possible platforms and approaches.
Light-matter coupling systems~\cite{Rabi-Braak,Eckle-Book-Models,JC-Larson2021} is an
ideal platform of quantum resources, while photon blockade~\cite{Boite2016-Photon-Blockade,Ridolfo2012-Photon-Blockade,LiaoJQ2020BlockadeJC,Ma2026PRL-Strong-2PB,Felicetti2026PRXQuantumBlockade, Carmichael1985,Birnbaum2005,Shamailov2010,Liew2010,Hamsen2017,Garziano2017,Flayac2017,Han2026WeakTPB}
is one of the clearest demonstrations for manipulation of light at the level
of individual quanta.

Indeed, with both the experimental progresses~\cite{Diaz2019RevModPhy,Kockum2019NRP,PRX-Xie-Anistropy,
Qin-ExpLightMatter-2018,WangYouJQ2023DeepStrong,Qin2024PhysRep,LiPengBo-Magnon-PRL-2024}
and the theoretical efforts~\cite{PRX-Xie-Anistropy,Braak2011,Boite2020},
light-matter coupling systems~\cite{Rabi-Braak,Eckle-Book-Models,JC-Larson2021} have
become an ideal platform for the explorations of quantum technologies with
high controllability and tunability. Light-matter interactions are also a fertile field for emerging quantum phenomenologies and novel quantum
theories~\cite{Braak2011,ChenQH2012,
Ashhab2013,Ying2015,LiuM2017PRL,Liu2021AQT,Hwang2015PRL,Hwang2016PRL,Irish2017,
Ying-g2hz-QFI-2024,*Ying-g2hz-QFI-2024-Cover,Ying-g1g2hz-QFI-2025,Ying2025g2A4,*Ying2025g2A4-Cover,Ying-g2Stark-QFI-2025,
Ying-2018-arxiv,Ying2020-nonlinear-bias,Ying-2021-AQT,*Ying-2021-AQT-Cover, Ying-gapped-top, Ying-Stark-top,*Ying-Stark-top-Cover,
Ying-Spin-Winding,*Ying-Spin-Winding-Cover,
Ying-JCwinding,Ying-Topo-JC-nonHermitian,*Ying-Topo-JC-nonHermitian-Cover,Ying-Topo-JC-nonHermitian-Fisher,*Ying-Topo-JC-nonHermitian-Fisher-Cover,Ying-gC-by-QFI-2024,
Grimaudo2022q2QPT,Grimaudo2023-Entropy,Grimaudo2024PRR,Zhu2024PRL,DeepStrong-JC-Huang-2024,PengJie2019,PengJ2021PRL,Padilla2022,Gao2022Rabi-dimer,GaoXL2025SPT,
Braak2019Symmetry,HiddenSymMangazeev2021,HiddenSymLi2021,HiddenSymBustos2021,
Irish2017,Irish-class-quan-corresp,
Felicetti2015-TwoPhotonProcess,e-collpase-Garbe-2017,e-collpase-Duan-2016,CongLei2019,Rico2020,
PengXHPRL2024RabiNMR,
Li2020conical,
KuangLM2024AQT,Ulstrong-JC-2,
Yan2023-AQT,ZhengHang2017,
Garbe2020,Montenegro2021-Metrology,Chu2021-Metrology,Garbe2021-Metrology,Ilias2022-Metrology, Ying2022-Metrology,YangZheng2023SciChina,Gietka2023PRL-Squeezing,Hotter2024-Metrology,Alushi2024PRL,Mukhopadhyay2024PRL,Mihailescuy2024,
Gietka2025PRL100802,Mihailescu2025CQMtutorial,QiuYi2025gA2,
Boite2016-Photon-Blockade,Ridolfo2012-Photon-Blockade,LiaoJQ2020BlockadeJC,Ma2026PRL-Strong-2PB,Felicetti2026PRXQuantumBlockade, Carmichael1985,Birnbaum2005,Shamailov2010,Liew2010,Hamsen2017,Garziano2017,Flayac2017,Han2026WeakTPB,QiaoFeng2026SpinSqueeze,QiaoFeng2016AsymPolaron},
with a wide relevance to quantum information and quantum computation~\cite{Diaz2019RevModPhy,Romero2012,Stassi2020QuComput,Stassi2018,Macri2018},
quantum metrology~\cite{Garbe2020,Montenegro2021-Metrology,Chu2021-Metrology,Garbe2021-Metrology,Ilias2022-Metrology, Ying2022-Metrology,Gietka2023PRL-Squeezing,YangZheng2023SciChina,Hotter2024-Metrology,Alushi2024PRL,Mukhopadhyay2024PRL,Mihailescuy2024, Ying-Topo-JC-nonHermitian-Fisher,*Ying-Topo-JC-nonHermitian-Fisher-Cover,Ying-g2hz-QFI-2024,*Ying-g2hz-QFI-2024-Cover, Ying-g1g2hz-QFI-2025,Ying2025g2A4,*Ying2025g2A4-Cover,Ying-g2Stark-QFI-2025,Gietka2025PRL100802,Mihailescu2025CQMtutorial,QiaoFeng2026SpinSqueeze,QiuYi2025gA2},
condensed matter~\cite{Kockum2019NRP}, nanowires~\cite{Nagasawa2013Rings,Ying2016Ellipse,Ying2017curvedSC,Ying2020PRR,Gentile2022NatElec}
and cold atoms~\cite{LinRashbaBECExp2013Review,LinRashbaBECExp2011,LiuYing02025exoticSOC2Ring,*LiuYing02025exoticSOC2Ring-Cover,LiuYing02025KaleidoscopeDDI}. In such a situation, photon blockade~\cite{Boite2016-Photon-Blockade,Ridolfo2012-Photon-Blockade,LiaoJQ2020BlockadeJC,Ma2026PRL-Strong-2PB,Felicetti2026PRXQuantumBlockade, Carmichael1985,Birnbaum2005,Shamailov2010,Liew2010,Hamsen2017,Garziano2017,Flayac2017,Han2026WeakTPB}
can be readily realized in light-matter coupling systems.

Photon blockade is a fundamental quantum optical phenomenon of photon
antibunching. Conventional photon blockade refers to one-photon blockade
(OPB)~\cite{Carmichael1985,Birnbaum2005}, whereas recently higher-order
photon blockades~\cite{LiaoJQ2020BlockadeJC,Shamailov2010,Hamsen2017},
especially two-photon blockade (TPB)~\cite{Hamsen2017,Ma2026PRL-Strong-2PB,Han2026WeakTPB,FengLJ2021TPB,Kudlaszyk2019PRATPB,QianBin2018praTPB,Miranowicz2013praTPB}, have been attracting more
and more attention. The OPB represents general photon antibunching
characterized by $g^{(n)}(0)<1$, where $g^{(n)}(0)$ ($n\geqslant 2$) are the
$n$th-order equal-time correlation functions, while TPB has two-photon
bunching ($g^{(2)}(0)>1$) but suppression of higher-order clustering ($g^{(3)}(0)<1$).
The OPB and higher-order blockade have been analyzed~\cite{Carmichael1985,Birnbaum2005,LiaoJQ2020BlockadeJC,Shamailov2010,Hamsen2017,Liew2010,Flayac2017}
in the Jaynes-Cummings (JC) regime~\cite{JC-model,JC-Larson2021} where the
coupling is weak and counter-rotating terms are negligible, while in the ultrastrong coupling regime the transition from OPB to TPB
has also been found~\cite{Ma2026PRL-Strong-2PB} and in deep-strong couplings a revival of photon blockade has been revealed~\cite{Boite2016-Photon-Blockade}. Conventional TPB (C-TPB) occurs with open
cascade decay chanel~\cite{Ma2026PRL-Strong-2PB,Han2026WeakTPB}. Recently an
unconventional TPB (U-TPB) was discovered in the situation of closed cascade
decay channel~\cite{Han2026WeakTPB}, emerging in the less-explored
intermediate coupling regime (strong-coupling). While the mechanism action
stage of the U-TPB is in the dissipation, the forming mechanism of U-TPB
lies in the interplay of weak anharmonicity and resonant driving~\cite{Han2026WeakTPB}.
The U-TPB implies both mechanism and manipulation diversities for the TPB.
However, this currently found U-TPB in linear coupling is limited in a
narrow coupling window and the blockade strength is weak, which would hinder
its applications. It is desirable to explore enhanced U-TPB with broadened
coupling regime and deepened blockade strength.

In this work, we propose to enhance the U-TPB by the Stark
nonlinear coupling~\cite{Eckle-2017JPA,*Eckle-2017JPA-b,Stark-Cong2020,Ying-Stark-top,*Ying-Stark-top-Cover}.
The introduction of the Stark nonlinear coupling to the linear
coupling enables tuning of both cascade energy level and the anharmonicity,
the former affects the C-TPB while the latter influences the U-TPB. By investigating photon
correlation functions and extracting phase diagrams in dissipation, we
demonstrate that our scheme not only dramatically broadens the window of
U-TPB but also deepens the
blockade degree. Indeed, the enhanced U-TPB can cover the entire strong coupling regime, and the U-TPB blockade strength can be upgraded by two orders.
The cascade decay rates, state
populations and the anharmonicity, are examined to identify and track the
U-TPB and C-TPB. In
the overview of phase diagrams, we also reveal a broad U-TPB phase
in the ultrastrong coupling regime, with a crossover to the U-TPB. Since the Stark nonlinear coupling is realizable
and tailorable, flexible both in the strength and in the sign~\cite{Stark-Grimsmo2013,Stark-Grimsmo2014,Stark-Cong2020,Zhai2025TwoPhotonStark},
our proposal may pave a practical way for manipulation of the TPB.

This paper is organized as follows.
Section~\ref{Section-Model} introduces
parity-broken generalized quantum Rabi model with Stark nonlinear coupling
in coherent driving.
Section~\ref{Section-Manipulation} gives an anylitical
analysis in the leading order to show the potential of manipulation via the
Stark coupling.
Section~\ref{Section-input-output} formulates the master
equation in dissipation and introduces input-output quantities in the
dressed picture.
Section~\ref{Section-enhanced-eg} illustrates an example of
U-TPB enhanced by the Stark coupling, with the key characters in the decay
rate, population, and weak anharmonicity.
Section~\ref{Section-diagram-enchance} shows the phase diagrams of enhancement.
Section~\ref{Section-diagram-enchance} shows the two enhancement branches and a broad U-TPB
phase in ultrastrong coupling, as well as the crossover to C-TPB.
Finally, Section~\ref{Section-Conclusion} summarizes our conclusions.

\section{Model}
\label{Section-Model}

The widely used model for the exploration of TPB is the generalized quantum
Rabi Hamiltonian describing a qubit coupled to a bosonic mode,
\begin{equation}
H_{\mathrm{Rabi}}=\omega a^{\dagger }a+\Omega \sigma ^{+}\sigma
^{-}+g(a+a^{\dagger })\bigl(\cos \theta \,\sigma _{z}-\sin \theta \,\sigma
_{x}\bigr),  \label{eq:H0}
\end{equation}%
where $a$ ($a^{\dagger }$) is the annihilation (creation) operator of the
bosonic mode, $\sigma ^{\pm }$ and $\sigma _{x,z}$ are Pauli operators of
the qubit, and $\theta $ is the coupling phase controlled by the external
bias. When $\theta =m\pi /2$, the Hamiltonian possesses the parity symmetry
as it commutes with the parity operator $\hat{P}=\sigma _{z}\exp (i\pi
a^{\dagger }a)$. For the value $\theta =0.3\pi $ used in this work, the
parity symmetry is broken, allowing dissipation decay that was previously
forbidden in parity symmetry to acquire non-zero decay transition matrix elements.

Note here that the coupling in $H_{\mathrm{Rabi}}$ with strength $g$ is linear, in the
sense that the coupling process is involving one photon emission and
absorption. In the present work we introduce a Stark coupling with the form~\cite{Eckle-2017JPA,*Eckle-2017JPA-b,Stark-Cong2020,Ying-Stark-top,*Ying-Stark-top-Cover}
\begin{equation}
H_{\mathrm{Stark}}=\chi \omega a^{\dagger }a\sigma _{z},  \label{H-Stark}
\end{equation}%
which has a nonlinear coupling strength in proportion to the photon number.
The Stark coupling can be realized and independently
tailored~\cite{Stark-Grimsmo2013,Stark-Grimsmo2014,Stark-Cong2020,Zhai2025TwoPhotonStark}.
The Stark coupling does not break the parity symmetry, thus favorable for
giving the vanishing zeroth order decay rate, which is crucial to maintain a suppressed cascacde decay channel.
So, in the present work, the system Hamiltonian
\begin{equation}
H_{0}=H_{\mathrm{Rabi}}+H_{\mathrm{Stark}}
\end{equation}%
is composed of both linear coupling and nonlinear coupling.

To investigate the input-output property, the system is driven by a weak
coherent field,
\begin{equation}
H_{\mathrm{drive}}=A_{d}\cos (\omega _{d}t)(a+a^{\dagger }),
\end{equation}%
with drive amplitude $A_{d}$ and frequency $\omega _{d}$. The total
Hamiltonian is then $H=H_{0}+H_{\mathrm{drive}}$. Throughout this work we set $%
\omega _{0}=\left( \omega +\Omega \right) /2$ and $\hbar =1$ as the units.

\begin{figure}[t!]
	\centering
	\includegraphics[width=0.80\linewidth]{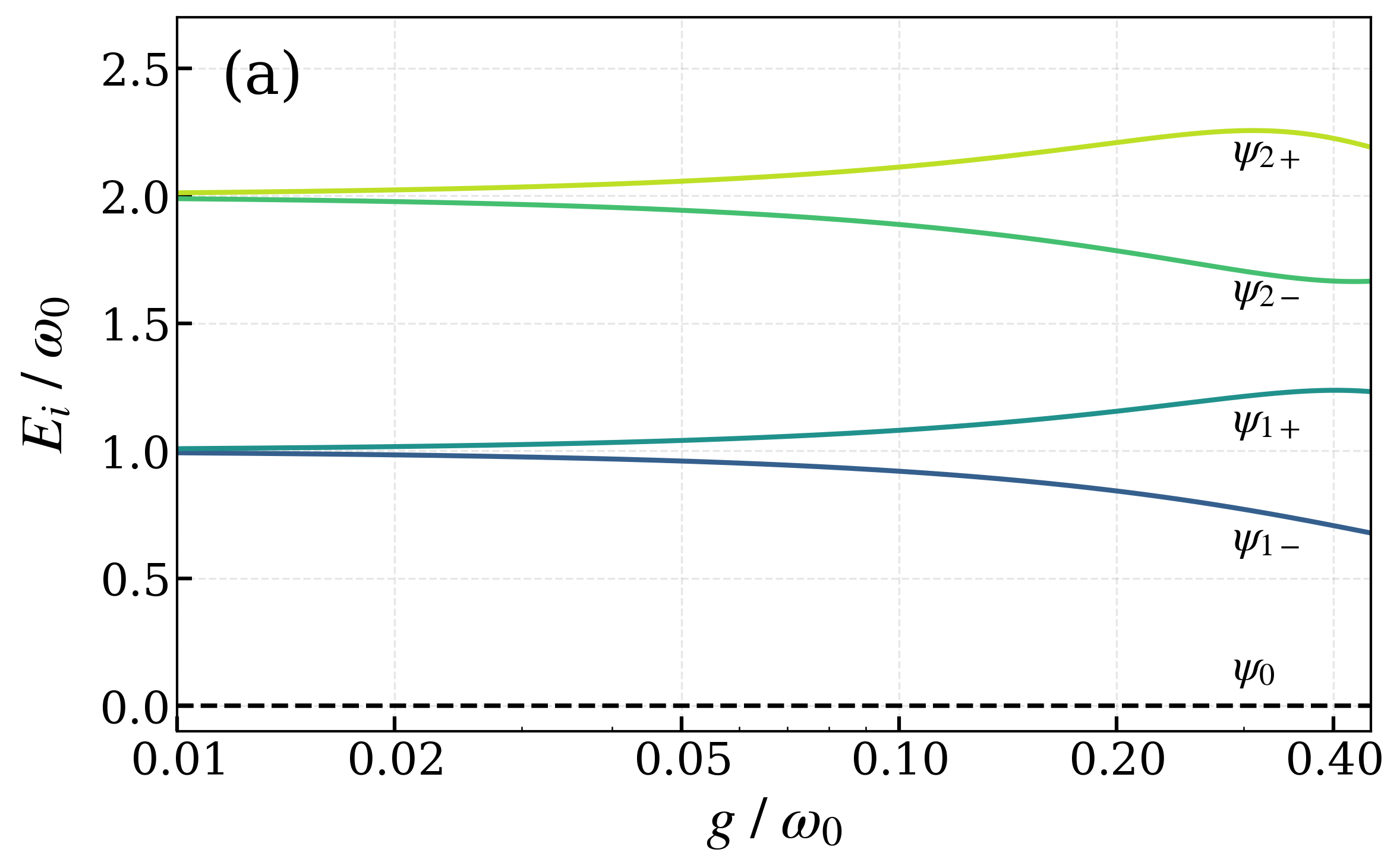}
	\includegraphics[width=0.80\linewidth]{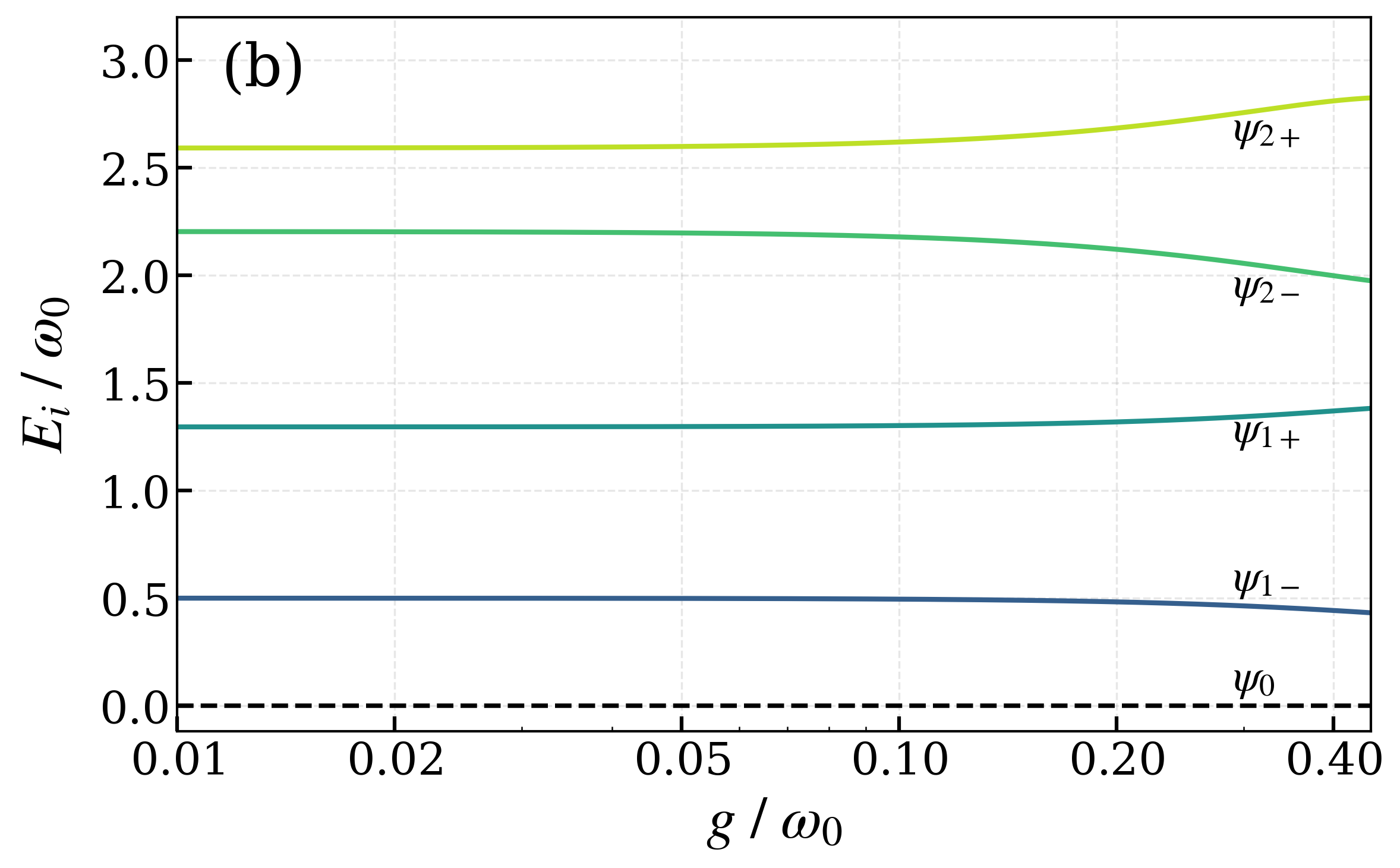}
	\caption{Potential of manipulation via the Stark coupling: Low energy levels $E_i$ of $H_0$, relevant for TPB,  versus linear coupling strength $g$ in the absence (a) and presence (b) of the Stark nonlinear coupling $\chi$. Here, the energy of the ground state $|\psi_0\rangle$ is set as the energy reference, $\omega_0=(\omega + \Omega)/2$ is set to be the energy unit and $\theta = 0.3\pi$. $\omega=1.0\omega_0$ and $\chi=0$ in (a); $\omega=1.5\omega_0$ and $\chi=0.136$ in (b).
}
\label{fig-Ej}
\end{figure}

\section{Potential of manipulation via the Stark coupling}
\label{Section-Manipulation}

\subsection{Energy level tuning in the Stark coupling}
\label{Section-Ei}

In Fig.\ref{fig-Ej} we illustrate some low energy levels of the system
Hamiltonian $H_{0}$, including the ground state $|\psi _{0}\rangle $ and two
sets of photon states which are relevant for the TPB studied in the present
work. The system will be coherently driven from the ground state $|\psi
_{0}\rangle $ onto the second excited state $|\psi _{1+}\rangle $ by tuning
the driving frequency to be the excitation energy of $|\psi _{1+}\rangle $,
while the fourth excited state $|\psi _{2+}\rangle $ will be involved in the
anharmonicity for the U-TPB.

As a comparison, Figure \ref{fig-Ej}(a) represents the case in the absence
of the Stark coupling while Figure \ref{fig-Ej}(b) denotes the case in the
presence of the Stark coupling. We see that the Stark coupling enables more
flexibility in tuning of both cascade energy level and the level above
cascade, which we will play key roles in the formation of TPBs. Indeed, the
gap between the primary cascade levels%
\begin{equation}
\Delta _{12}=E_{1,+}-E_{1,-}  \label{Eq-gap12}
\end{equation}%
and the anharmonicity parameter%
\begin{equation}
\Delta =(E_{2,+}-E_{1,+})-(E_{1,+}-E_{0})  \label{Eq-anharmon}
\end{equation}%
are important parameters in deciding the properties of the TPB. We can
roughly see the role of Stark coupling in tuning the gap and anharmonicity
in the JC\ regime, where the counter-rotating terms are neglected. If we
also regard the coupling in the $\sigma _{z}$ direction as a perturbation
due to small $g$, the energy in the leading order is then explicitly
available as~\cite{Ying-Stark-top,Ying-JCwinding}%
\begin{eqnarray}
E_{n,\pm }^{\mathrm{JC-Stark}} &=&e_{+}\pm \sqrt{e_{-}^{2}+n\ g_{x}^{2}}, \\
E_{0}^{\mathrm{JC-Stark}} &=&-\frac{\Omega }{2},
\end{eqnarray}%
with $g_{x}=-g\sin \theta $, $e_{+}=\left( n-\frac{1+\chi }{2}\right) \omega
$ and $e_{-}=\frac{1}{2}\left( \Omega -\omega \right) +\left( n-\frac{1}{2}%
\right) \chi \omega $. The corresponding eigen-wavefunctions are%
\begin{eqnarray}
\psi _{n,\pm }^{\mathrm{JC-Stark}} &=&\left( C_{n\uparrow }^{\left( \pm
\right) }\left\vert n-1,\uparrow \right\rangle +C_{n\downarrow }^{\left( \pm
\right) }\left\vert n,\downarrow \right\rangle \right) /\sqrt{N_{n,\pm }},
\label{WaveF-JC-n} \\
\psi _{0}^{\mathrm{JC-Stark}} &=&\left\vert 0,\downarrow \right\rangle ,
\end{eqnarray}%
where $n=1,2,\cdots $ and $\uparrow ,\downarrow $ are spin states of $\sigma
_{z}$. The basis coefficients are given by $C_{n\uparrow }^{\left( \pm
\right) }=e_{-}\pm \sqrt{e_{-}^{2}+n\ g_{x}^{2}},C_{n\downarrow }^{\left(
\pm \right) }=g_{x}\sqrt{n}$and $N_{n,\pm }=C_{n\Uparrow }^{\left( \pm
\right) 2}+C_{n\Downarrow }^{\left( \pm \right) 2}$ is the normalization
factor.

Beyond the JC regime, the perturbation from the counter-rotating terms and
the $\sigma _{z}$ coupling component $\delta H=g\cos \theta (a+a^{\dagger
})\,\sigma _{z}-g\sin \theta (a\sigma _{-}+a_{--}^{\dagger }\sigma _{+})\,$\
yields an $O(g^{2})$ correction to the energy%
\begin{equation}
E_{j}=E_{j}^{\mathrm{JC-Stark}}+\sum_{k\neq j}\frac{\left\vert \left\langle
\psi _{k}^{\mathrm{JC-Stark}}\left\vert \delta H\right\vert \psi _{j}^{%
\mathrm{JC-Stark}}\right\rangle \right\vert ^{2}}{E_{j}^{\mathrm{JC-Stark}%
}-E_{k}^{\mathrm{JC-Stark}}}  \label{Eq-Perturbation-Ej}
\end{equation}%
where $j=0,n-,n+.$ Here the first-order perturbation $\left\langle \psi
_{j}^{\mathrm{JC-Stark}}\left\vert \delta H\right\vert \psi _{j}^{\mathrm{
JC-Stark}}\right\rangle $ vanishes due to the broken symmetry of excitation
number by $\delta H$. The second order correction is of quadratic order $\mathcal{O}(g^{2}).$

We see that the gap in this\ regime%
\begin{equation}
\Delta _{12}=\sqrt{\left[ \left( \Omega -\omega \right) +\chi \omega \right]
^{2}+4g_{x}^{2}}+C_{12}^{\mathrm{pert}}g^{2}
\end{equation}
is tunable by the Stark coupling parameter $\chi $. On the other hand, the
anharmonicity in the leading orders takes the form%
\begin{equation}
\Delta =C_{0}\left( \omega -\Omega \right) +C_{1}\chi \omega
+C_{2}g_{x}^{2}+C^{\mathrm{pert}}g^{2}
\end{equation}
where $C_{0}=(1+2s_{1}-s_{2})/2$, $C_{1}=(1-2s_{1}+3s_{2})/2$,
$C_{2}=[2(s_{1}-s_{2})(\Omega -\omega )+2(s_{1}-3s_{2})\chi \omega
]/(s_{1}s_{2}e_{1}e_{2})$
and
$s_{1},s_{2}$ are the signs of $e_{1}=\left(
\Omega -\omega \right) +\chi \omega $ and $e_{2}=\left( \Omega -\omega
\right) +3\chi \omega $. Here $C_{12}^{\mathrm{pert}}$
and
$C^{\mathrm{pert}}$ are perturbation coefficients from the second-order correction in (\ref{Eq-Perturbation-Ej}).
Apparently, the anharmonicity is also adjustable by
the Stark coupling. Such a level tunability provides the possibility of
manipulation of the TPB and leads to enhancement of the U-TPB, as we will
address below.

\subsection{Release of detuning dimension}

The U-TPB occurs in the presence of weak anharmonicity~\cite{Han2026WeakTPB}
with small values of $\Delta $, which requires the vicinity of the resonance
situation $\omega =\Omega $. Now with the Stark coupling, we can go away
from resonance and release the detuning dimension with unequal cavity
frequency $\omega $ and qubit splitting energy $\Omega $. Indeed, as we will
see, the combination of the detuning and the Stark coupling will help to
optimize the enhancement of U-TPB.

\subsection{Symmetry of Stark coupling favorable for maintaining the suppression of decay channel}

The Stark coupling does not break the symmetry of parity or the excitation number,
\begin{equation}
\hat{N}=a^{\dagger }a+\sigma _{z}/2+1/2
\end{equation}
whose eigenvalue $N$ is also equal to the photon number label in $\psi _{n,\pm }^{\mathrm{JC-Stark}}$. This symmetry feature of the Stark coupling is favorable for maintaining the suppression of cascade decay channel which is the character of the U-TPB. Indeed, as will be seen in Eq.~(\ref{Eq-Decay-Rate}), the transition rate of the primary cascade decay channel $\Gamma _{1+,1-}$ is proportional to the square of the transition matrix element
$C_{1-,1+}^{c}=\left\langle \psi _{1-}\left\vert X^{c}\right\vert \psi _{1+}\right\rangle$
where
$X^{c}=c-c^{\dagger }$ and $c=a,\sigma ^{-}$.
The vanishing of the zeroth order of the transition matrix element is a necessary condition for the formation of the U-TPB. The Stark coupling rightly meets this condition as
the zeroth-order matrix element vanishes exactly
\begin{equation}
\left( C_{1-,1+}^{c} \right) _{0}
=\left\langle    \psi^{\mathrm{JC-Stark}} _{1-}  \left\vert X^{c} \right\vert   \psi^{\mathrm{JC-Stark}} _{1+}   \right\rangle
=0,  \label{Eq-C11-0th}
\end{equation}
due to that $X^c$ breaks the symmetry of
excitation number and induces transition from the $\psi _{1+}$ state in the $N=1$ sector to states in other $N$ sectors which are orthogonal to the $N=1$
sector. Then, non-vanishing contribution lies in higher orders of the transition matrix element that will be at least linear or quadratic order of the coupling $g$.
With the square form of the transition matrix element in $\Gamma _{1+,1-}$, the decay transition rate will be no less than quadratic order. Indeed, the perturbation analysis shows that the qubit decay and cavity dissipation respectively lead to quadratic ($g^2$) and quartic ($g^4$) contributions to the decay transition rate in the leading order. In such a situation, the suppression of the cascade decay channel is very likely to happen.

These characters of manipulation potential and symmetry preservation render the Stark coupling to be a favorable control parameter for the enhancement of the U-TPB.

\section{Input-output quantities in dissipation}

\label{Section-input-output}

To correctly describe dissipation properties in all coupling regimes, including the weak, strong and ultrastrong ones, we
work in the dressed picture~\cite{Ridolfo2012-Photon-Blockade,RidolfoPRL2013,Felix2011DressedPicture,Nori2018PRADressed}. The system Hamiltonian $H_{0}$ is exactly
diagonalized~\cite{Ying2020-nonlinear-bias,
Ying-Stark-top,*Ying-Stark-top-Cover,
Ying-g1g2hz-QFI-2025} in the Fock space to obtain the dressed eigenstates $|\psi
_{j}\rangle $ and eigenenergies $E_{j}$ with the eigenequation $H_{0}|\psi
_{j}\rangle =E_{j}|\psi _{j}\rangle $. We label the ground state by $|\psi
_{0}\rangle $, the two branches of states by $|\psi _{n\pm }\rangle $
corresponding to the two branches of $n$-photon states $|\psi _{n\pm }^{%
\mathrm{JC-Stark}}\rangle $ in JC regime~\cite{Ying-Stark-top,Ying-JCwinding}. The transition frequencies are
denoted as $\Delta _{jk}=E_{j}-E_{k}$.

The Lindblad master equation that governs the dynamics in dissipation reads
\begin{equation}
\dot{\rho}=-i[H_{\mathrm{rot}},\rho ]+\sum_{c=a,\sigma ^{-}}\mathcal{L}%
_{c}[\rho ]+\mathcal{L}_{deph}[\rho ],
\end{equation}%
where $\mathcal{D}[L]\rho $ is the standard Lindblad dissipatior in the
dressed basis. The second term represents the dissipator
\begin{equation}
\mathcal{L}_{c}[\rho ]=\sum_{j>k}\Gamma _{jk}^{c}\mathcal{D}\bigl[|\psi
_{k}\rangle \langle \psi _{j}|\bigr]\rho ,
\end{equation}%
with the transition rates
\begin{equation}
\Gamma _{jk}^{c}=\gamma _{c}\frac{|\Delta _{jk}|}{\omega _0 }%
|C_{jk}^{(c)}|^{2}.  \label{Eq-Decay-Rate}
\end{equation}%
Here, $\gamma _{c}$ is the bare decay rate of operator $c$ and the matrix
elements are defined as
\begin{equation}
C_{jk}^{(c)}=-i\langle \psi _{j}|(c-c^{\dagger })|\psi _{k}\rangle \quad
(c=a,\sigma ^{-}).  \label{Eq-Cjk}
\end{equation}%
The last term in the master equation denotes the qubit dephasing
\begin{equation}
\mathcal{L}_{\mathrm{deph}}[\rho ]=\sum_{j}\Gamma _{j}^{\mathrm{deph}}%
\mathcal{D}[|j\rangle \langle j|]\rho ,
\end{equation}%
where $\Gamma _{j}^{\mathrm{deph}}=\gamma _{\mathrm{deph}}\big|\langle
j|\sigma _{z}|j\rangle \big|$.

\begin{figure}[t!]
	\centering
	\includegraphics[width=0.88\linewidth]{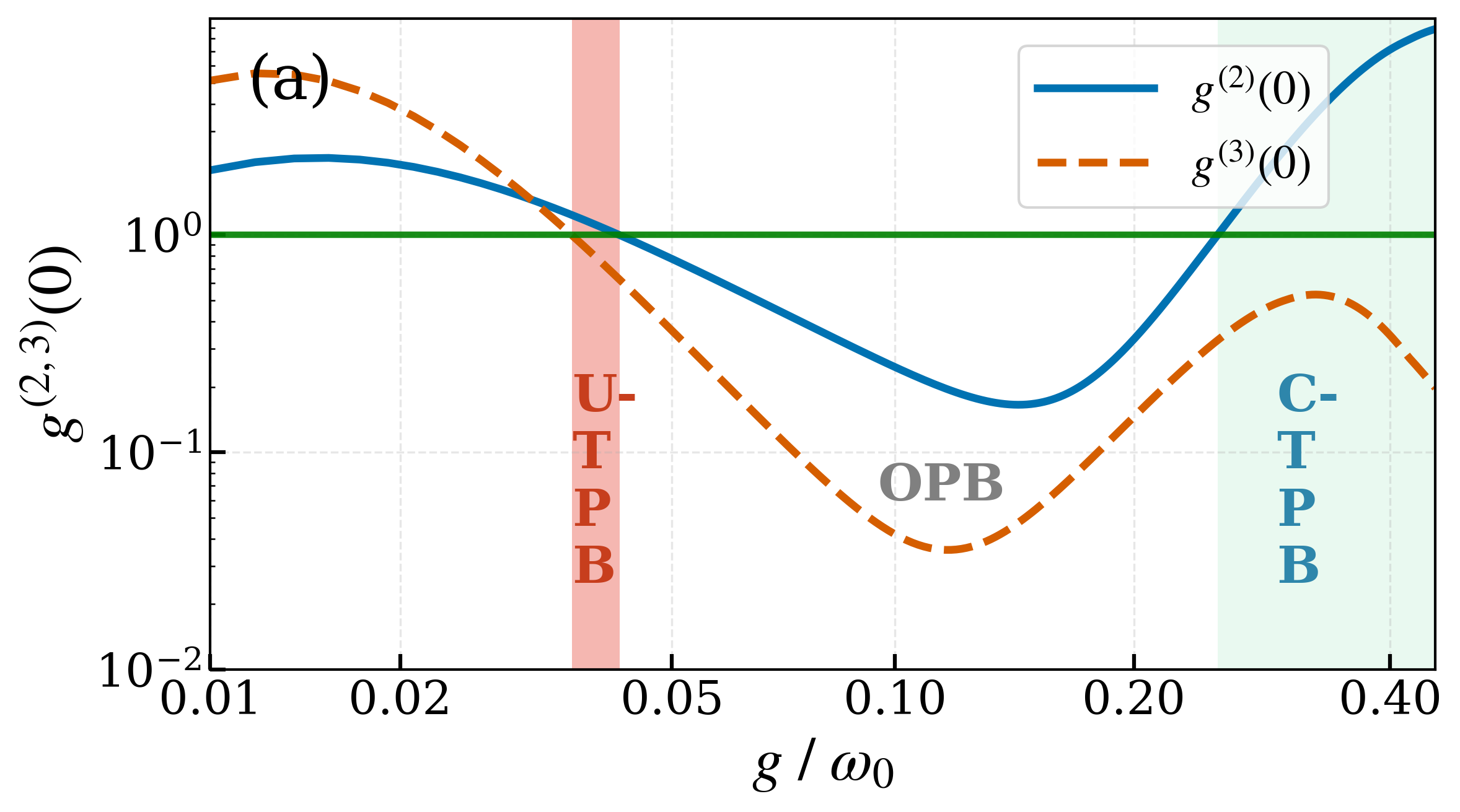}
    \includegraphics[width=0.88\linewidth]{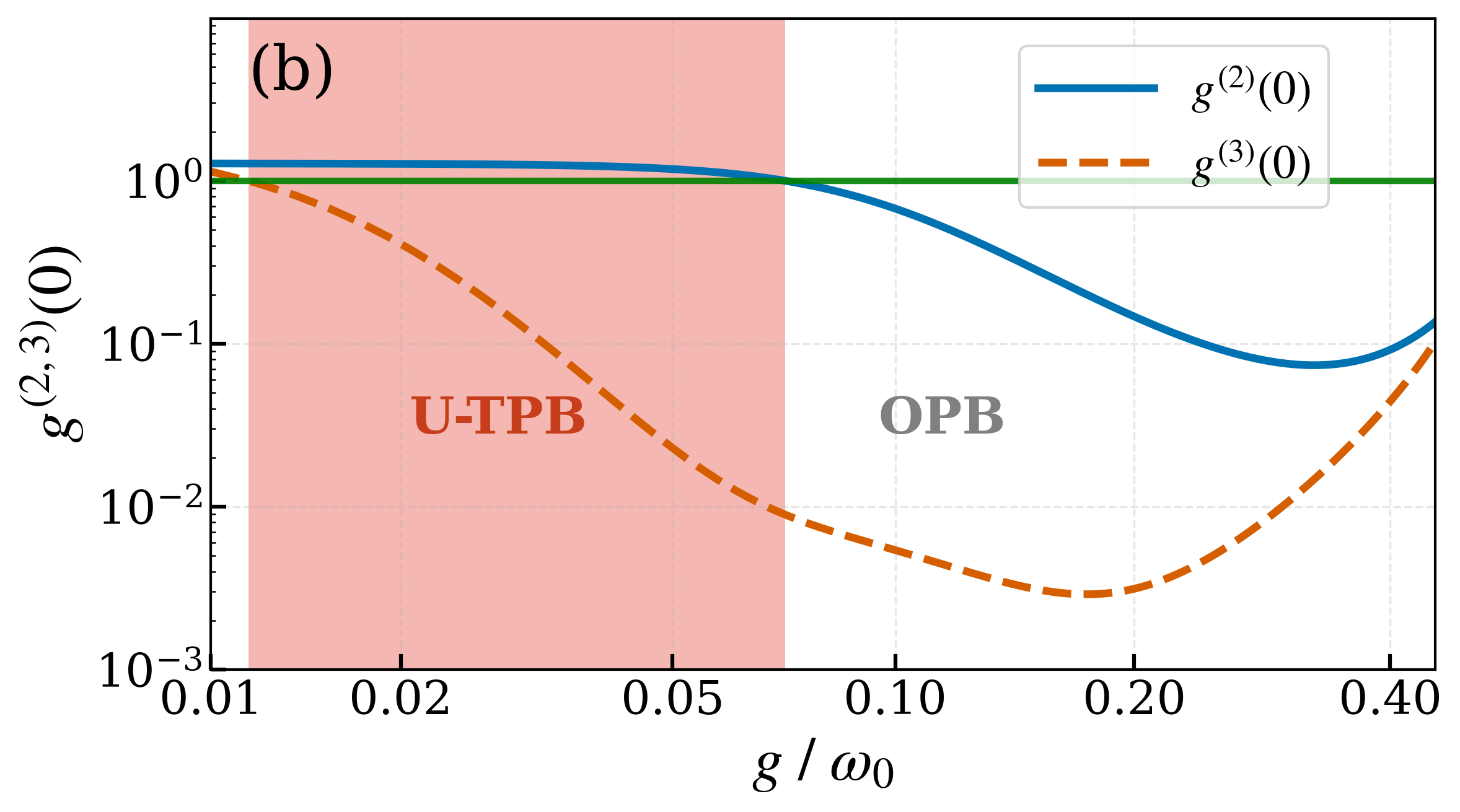}
	\caption{Enhanced unconventional two-photon blockade (U-TPB) without primary cascade channel by the Stark coupling $\chi$ in strong linear coupling regime ($0.01\lesssim g/\omega_0\lesssim 0.1$), apart from the conventional TPB (C-TPB) with primary cascade channel in ultrastrong-coupling (USC, $g/\omega_0\gtrsim 0.1$) regime. TPB is identified by $g^{(2)}(0) > 1$ and $g^{(3)}(0) < 1$ under coherent driving from the ground state $|\psi_0\rangle$ onto the $|\psi_{1+}\rangle$ state, where $g^{(2)}(0)$ (solid) and $g^{(3)}(0)$ (dashed) are the second- and third-order equal-time correlation functions.
(a) $g^{(2)}(0)$ and $g^{(3)}(0)$ versus $g$ at $\chi=0.0$.
(b) $g^{(2)}(0)$ and $g^{(3)}(0)$ versus $g$ at $\chi=0.136$.
In (a) and (b), the pale-rose (gray) shaded region marks the U-TPB window, while C-TPB is in the light-minty-green region. The closing or opening of primary cascade decay channel is shown in Fig.~\ref{Fig-dacay-rates}. Here, we set the diving strength $A_d = 10^{-3}\omega_0 $ and dissipation factors $\gamma_a = \gamma_{\sigma^-} =2 \gamma_{\rm deph}=10^{-2}\omega_0  $, while other system parameters are the same as in Fig.~\ref{fig-Ej}. Unless otherwise specified, the parameters are the same in other figures.
}
\label{Fig-g2g3-enhanced}
\end{figure}
\begin{figure}[h!]
	\centering
    \includegraphics[width=0.88\linewidth]{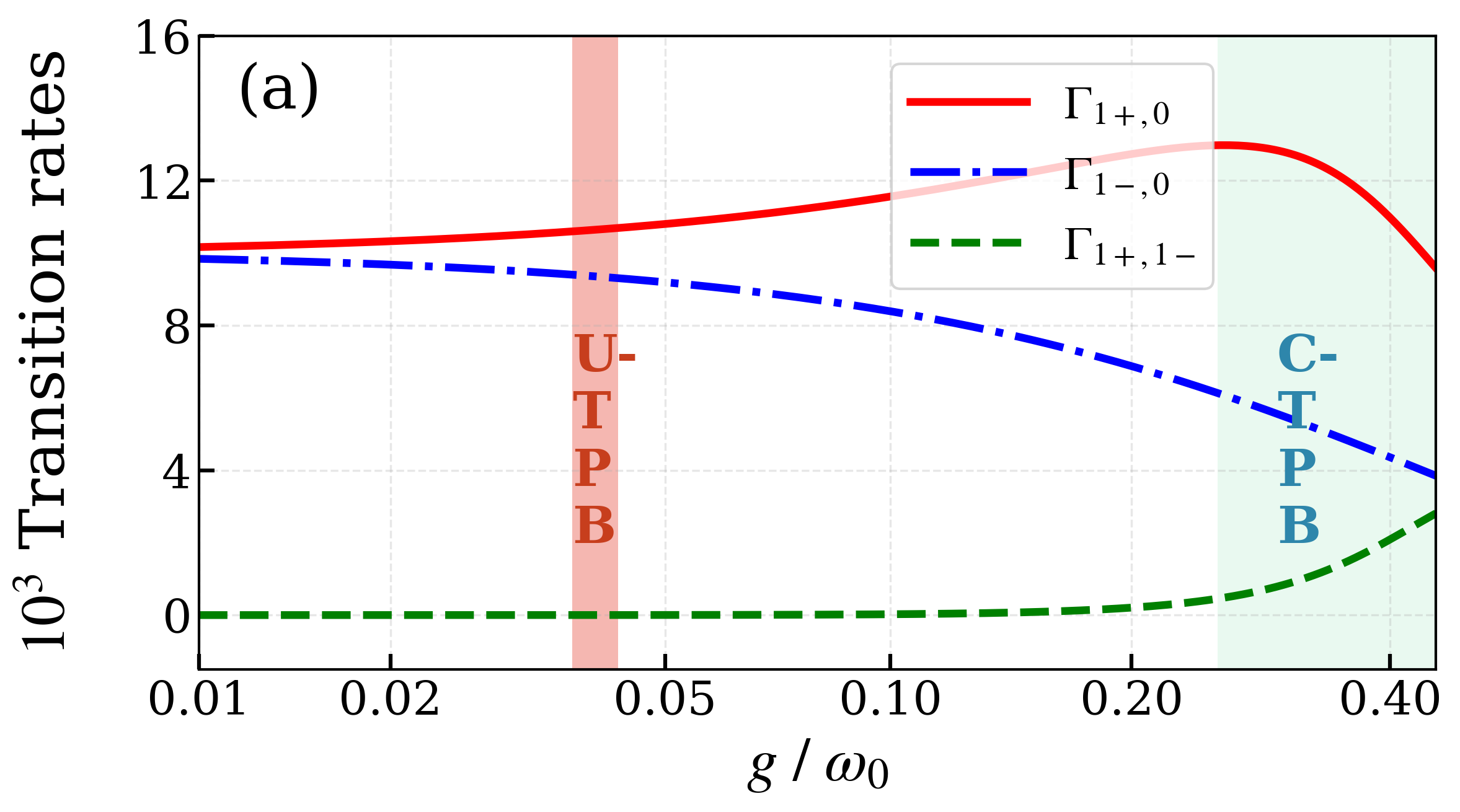}
    \includegraphics[width=0.88\linewidth]{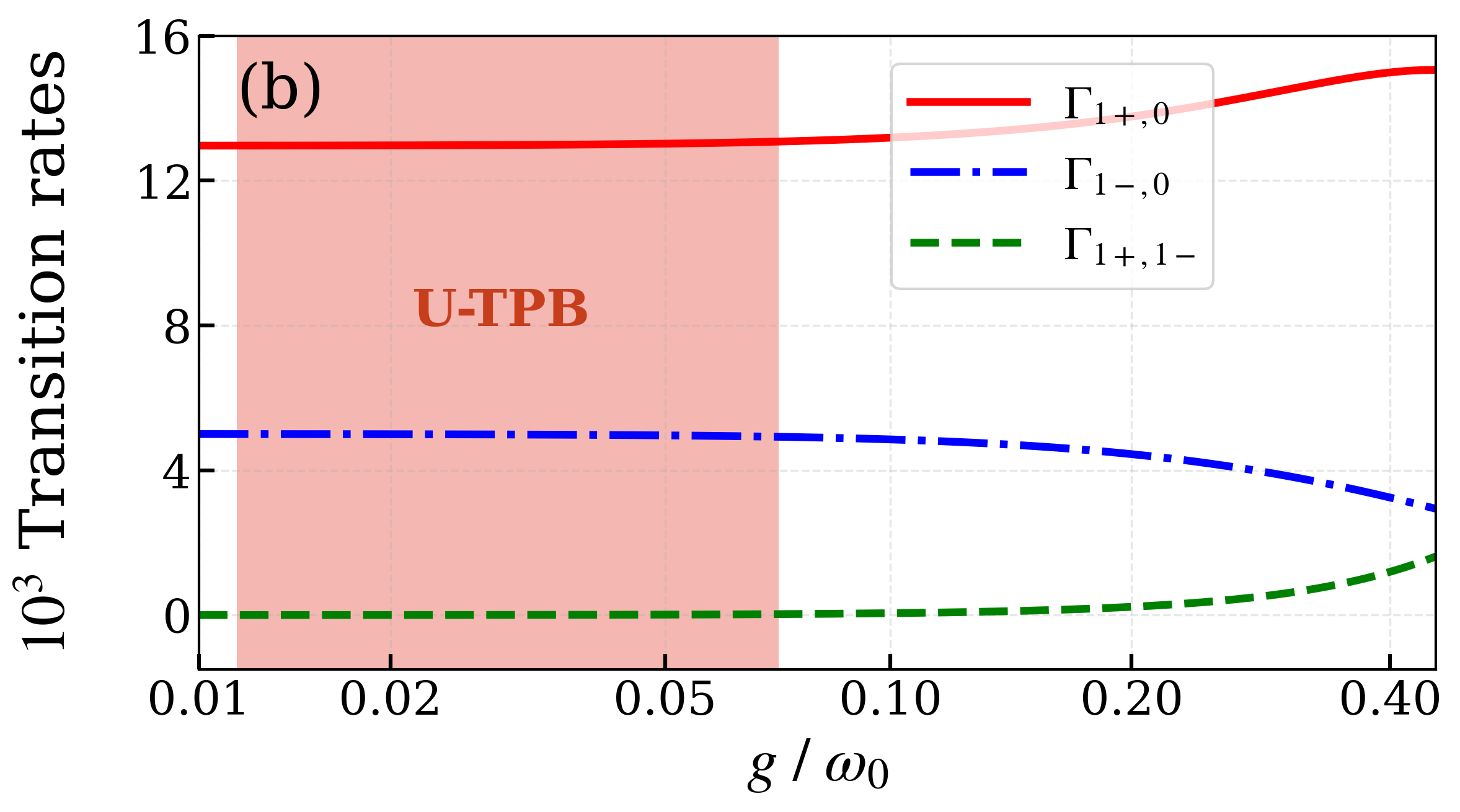}
	\caption{Transition rates of different decay channels versus $g/\omega_0$: $\Gamma_{1+,0}$ (solid red), $\Gamma_{1-,0}$ (dash-dotted blue), and $\Gamma_{1+,1-}$ (dashed green): (a) $\chi=0.0$, (b) $\chi=0.136$. The primary cascade channel $\Gamma_{1+,1-}$ remains negligible in the U-TPB regime, while it becomes large in the C-TPB regime.  }
\label{Fig-dacay-rates}
\end{figure}

The output field operator is related to the input field by
\begin{equation}
a_{\mathrm{out}}(t)=a_{\mathrm{in}}(t)-i\Upsilon \dot{X}^{+}(t),
\label{Eq-input-ouput}
\end{equation}%
which is valid for all couplings \cite{Ridolfo2012-Photon-Blockade}. Here $%
\Upsilon $ is the waveguide coupling coefficient and $\dot{X}^{+}$ is the
positive-frequency component of the time derivative of the operator $%
X=-i(a-a^{\dagger })$, expressed in the dressed-state basis:
\begin{equation}
\dot{X}^{+}=-i\sum_{j>k}\Delta _{kj}\,X_{jk}\,|\psi _{j}\rangle \langle \psi
_{k}|,
\end{equation}%
with $X_{jk}=\langle \psi _{j}|X|\psi _{k}\rangle $.

The $n$th-order equal-time correlation functions of the output field are
then defined as
\begin{equation}
g^{(n)}(0)=\frac{\langle (\dot{X}^{-})^{n}(\dot{X}^{+})^{n}\rangle }{\langle
\dot{X}^{-}\dot{X}^{+}\rangle ^{n}},  \label{eq:gn}
\end{equation}%
where $\dot{X}^{-}$ is the Hermitian conjugate of $\dot{X}^{+}$. This
definition properly incorporates both photonic and atomic contributions to
the output field statistics in the presence of strong counter-rotating terms
and parity breaking.

\section{Enhancement of U-TPB}

\label{Section-enhanced-eg}

In this Section we will illustrate that in the presence of Stark coupling
the U-TPB can be enhanced both with a broadened coupling window and with a
deepened blockading degree. The characters of U-TPB remain in the enhanced
U-TPB, including the closed cascade channel and weak anharmonicity scenario.

\subsection{Narrow window of U-TPB in the absence of Stark coupling}

To identify possible signatures of TPB, we calculate the second- and
third-order equal-time correlation functions $g^{(2)}(0)$ and $g^{(3)}(0)$,
defined in Eq.~(\ref{eq:gn}), as functions of the coupling strength $g$. The
system is driven resonantly at the upper-photon frequency $\omega
_{d}=E_{1+}-E_{0}$, with the system parameters $\theta =0.3\pi $, $\Omega
=10^{-3}\omega _{0}$, $\gamma _{a}=\gamma _{\sigma ^{-}}=2\gamma _{\mathrm{%
deph}}=10^{-2}\omega _{0}$~\cite{Ma2026PRL-Strong-2PB,Han2026WeakTPB}. As
mentioned in Introduction, the occurrence of one-photon blockade (OPB) is
judged by correlation functions reduced from unity $g^{(2)}(0)<1$ and $%
g^{(3)}(0)<1$, while the TPB is identified by $g^{(2)}(0)>1$ and $%
g^{(3)}(0)<1$~\cite%
{Boite2016-Photon-Blockade,Ridolfo2012-Photon-Blockade,LiaoJQ2020BlockadeJC,Ma2026PRL-Strong-2PB,Carmichael1985,Birnbaum2005,Shamailov2010,Liew2010,Hamsen2017,Garziano2017,Flayac2017}%
.

In the absence of Stark coupling, Figure \ref{Fig-g2g3-enhanced}(a) shows
the U-TPB (pale-rose region) found in a strong-coupling regime apart from
the C-TPB (light-minty-green) in the ultrastrong coupling regime. The regime
before the U-TPB is non-blockading ($g^{(2)}(0)>1$ and $g^{(3)}(0)>1$),
while emerging in the regime between the U-TPB and C-TPB is a one-photon blockade (OTB)
phase.

The U-TPB opens an avenue for creating TPB without opening the cascade channel differently from the C-TPB,
as we see in Fig~\ref{Fig-dacay-rates}(a) and discussed in more detail in Section~\ref{Section-decay-rate}.
However, we notice that the U-TPB phase is in a narrow coupling window and the
blockading degree is weak as the amplitude of $g^{(3)}(0)$ is not much
smaller than $1$. It would be favorable to have an enhanced U-TPB with wider
phase and stronger blockading degree.

\subsection{Broadened window of U-TPB in the presence of Stark coupling}

We find the enhancement of the U-TPB can be realized in the presence of the Stark
coupling. Indeed, as illustrated in Figure \ref{Fig-g2g3-enhanced}(b) with a
finite Stark coupling $\chi =0.136$, we find an U-TPB phase (also pale-rose
region) which spans over a much broadened coupling regime from $g\simeq
0.01\omega _{0}$ to $g\simeq 0.07\omega _{0}$, in a sharp contrast to the
narrow window $0.037\lesssim g/\omega _{0}\lesssim 0.043$ in Figure \ref{Fig-g2g3-enhanced}(a).

\subsection{Deepened blockading degree of U-TPB in the presence of Stark
coupling}

We also find that the blockading degree of the U-TPB is also deepened in the
presence of Stark coupling. Indeed, in the presence of Stark coupling the
value of $g^{(3)}(0)$ can be dramatically reduced, up to two orders smaller
than that in the absence of Stark coupling, as we see in Fig.~\ref{Fig-g2g3-enhanced}(b)
in comparison with Fig.~\ref{Fig-g2g3-enhanced}(a).
These results illustrate that the Stark coupling can effectively enhance the
U-TPB both in the coupling width and the blockading degree, which may actualize from a
limited conceptional phenomenon to be more practical for applications.

\begin{figure}[t!]
	\centering
    \includegraphics[width=0.88\linewidth]{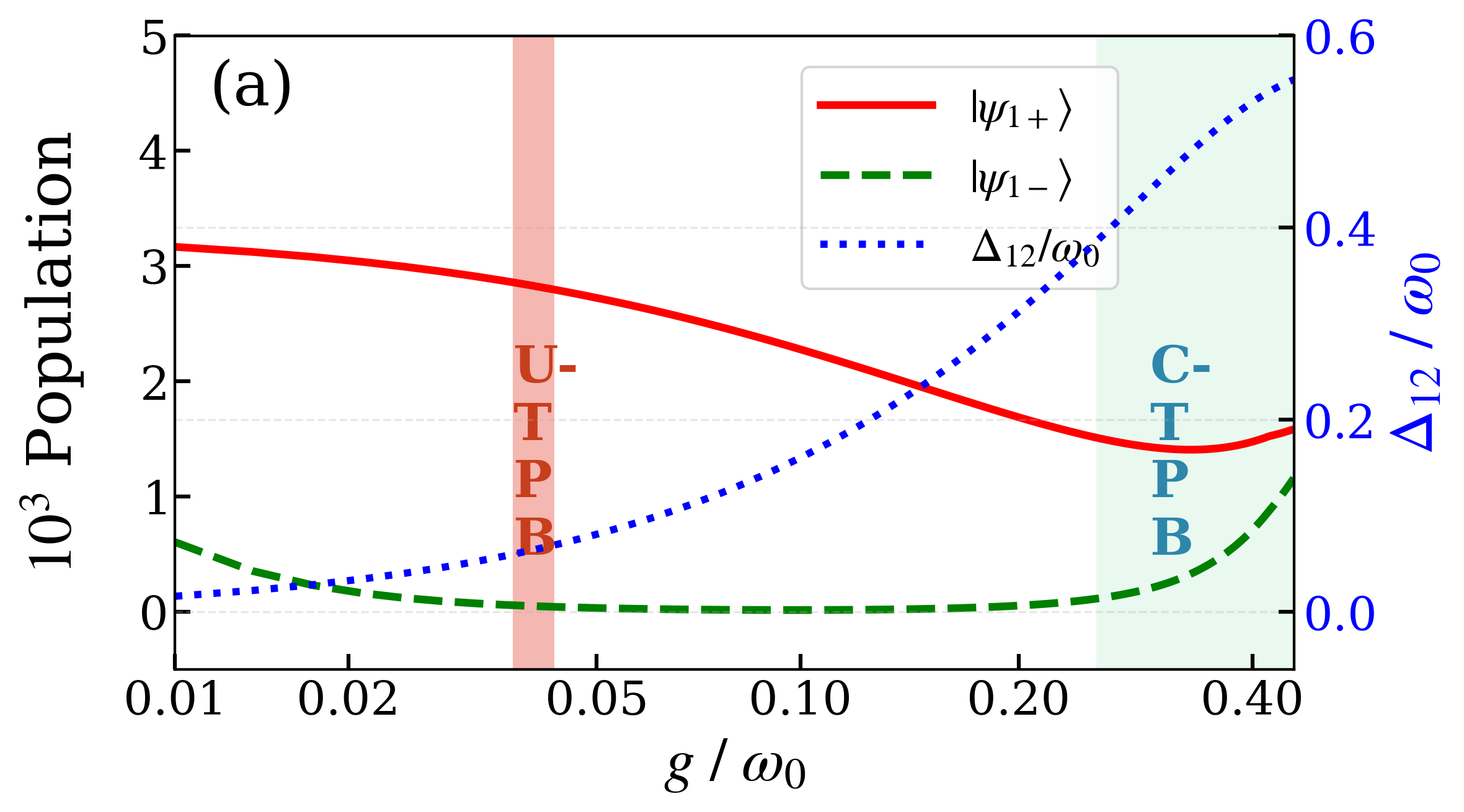}
    \includegraphics[width=0.88\linewidth]{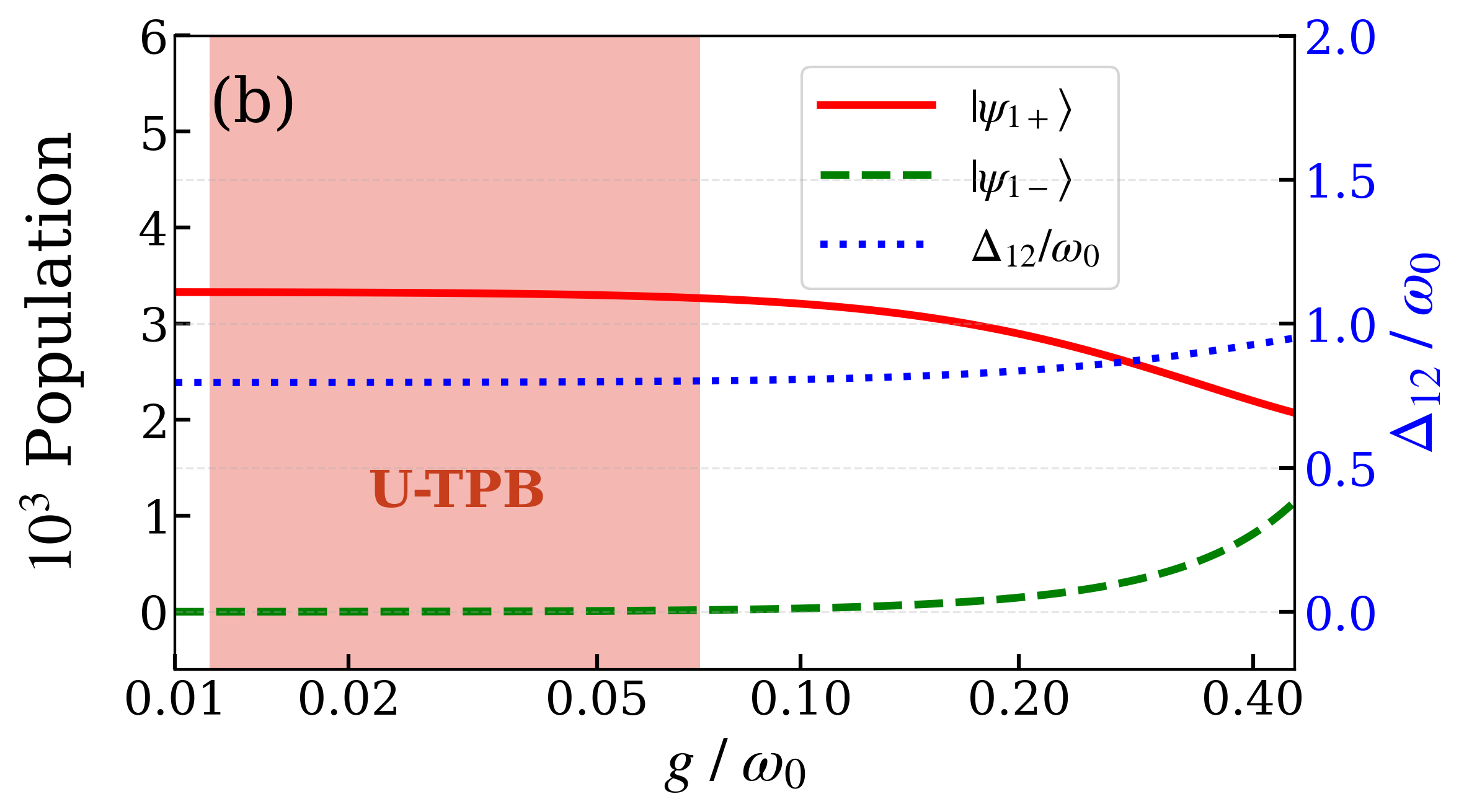}
	\caption{Populations (left axis) of states $|\psi _{1+}\rangle $ (solid line) and $|\psi _{1-}\rangle $ (dashed line) in cascade decay and the gap $\Delta_{12}$ (dotted line, right axis) between these two state: (a) $\chi=0.0$, (b) $\chi=0.136$. The $|\psi _{1-}\rangle $ population is vanishing in the U-TPB while it is finite in the C-TPB. }
\label{Fig-population}
\end{figure}

\begin{figure}[thbp]
	\centering
    \includegraphics[width=0.88\linewidth]{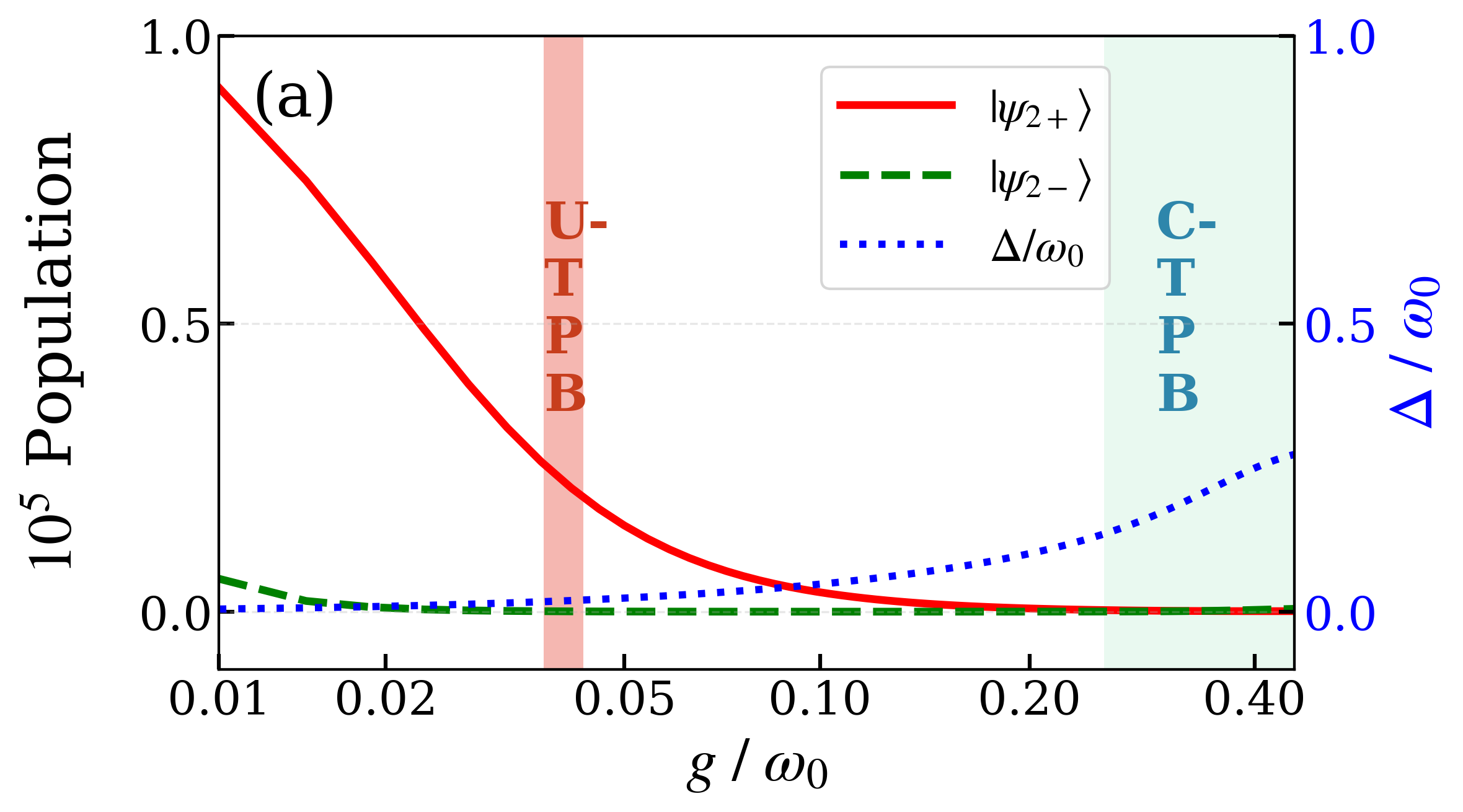}
    \includegraphics[width=0.88\linewidth]{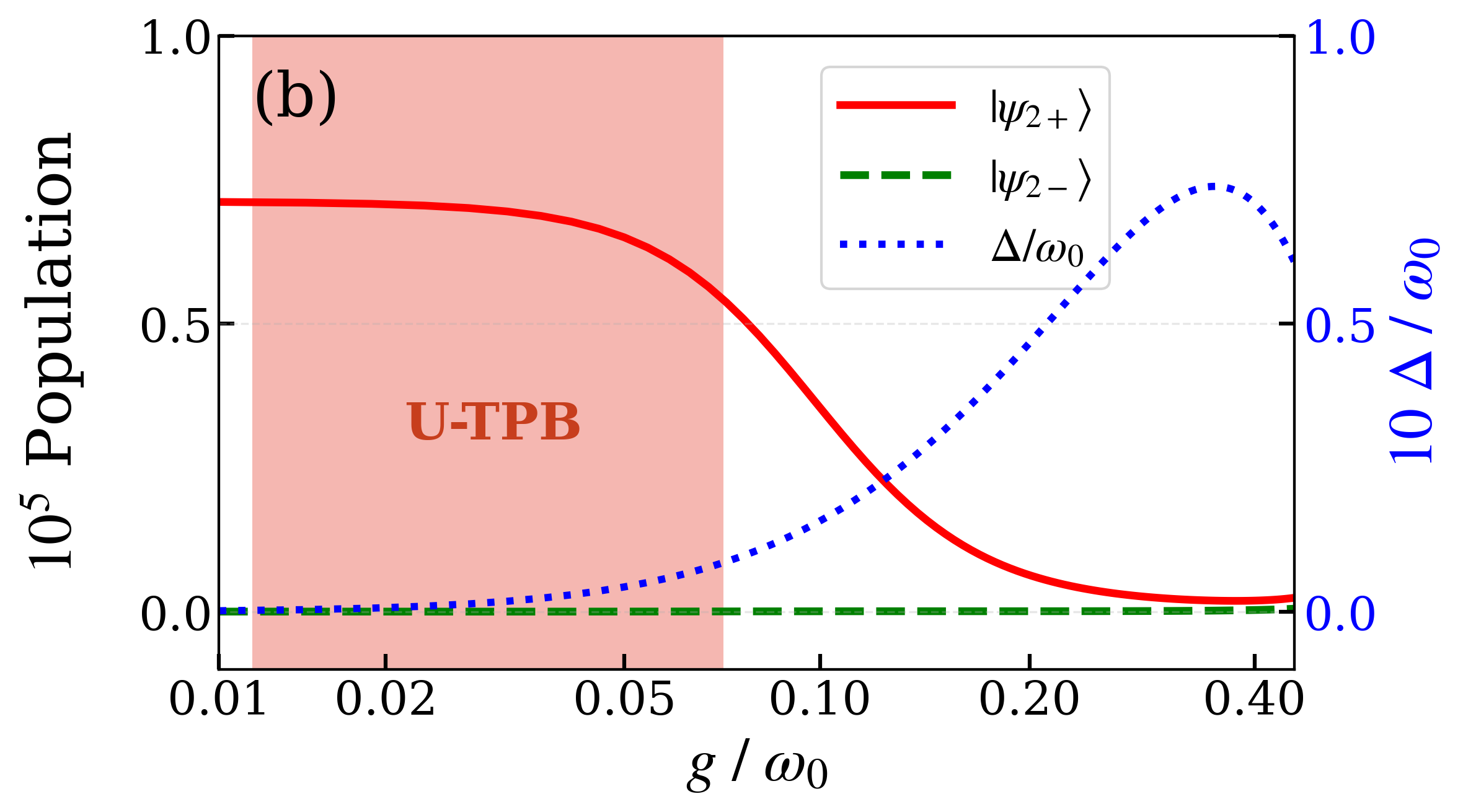}
	\caption{Weak anharmonicity (dotted line, right axis) and populations (lrft axis) of states $|\psi _{2+}\rangle $ (solid line) and $|\psi _{2-}\rangle $ (dashed line) above decay cascade: (a) $\chi=0.0$, (b) $\chi=0.136$. The $|\psi _{2+}\rangle $ population is non-vanishing due to resonance effect in weak anharmonicity, which enables the formation of the U-TPB without the conventional cascade decay channel of the C-TPB in Fig.~\ref{Fig-population}.}
\label{Fig-anharmonicity}
\end{figure}

\subsection{Identifying the enhanced U-TPB: Vanishing transition rate and
population in the cascade decay}\label{Section-decay-rate}

The U-TPB here is distinguished from the C-TPB by the closed primary cascade decay
channel. When the system is driven from the ground state $|\psi _{0}\rangle $
to the second excited state $|\psi _{1+}\rangle $, as in the standard study
of TPB, there are three cascade decay channels: $|\psi _{1+}\rangle
\rightarrow |\psi _{0}\rangle $, $\quad |\psi _{1-}\rangle \rightarrow |\psi
_{0}\rangle $, and $|\psi _{1+}\rangle \rightarrow |\psi _{1-}\rangle $.
Among them, the closure of the primary channel $|\psi _{1+}\rangle \rightarrow |\psi
_{1-}\rangle $ would means a OPB while its opening signals the transition to
the C-TPB as in the conventional TPB picture.

Figure~\ref{Fig-dacay-rates}(a) shows the transition rates $\Gamma _{1+,0}$
(solid red), $\Gamma _{1-,0}$ (dash-dotted blue), and $\Gamma _{1+,1-}$
(dashed green), with their vanishing or finite values meaning the closure or
opening of the afore-mentioned cascade channels respectively. We see in Fig.~%
\ref{Fig-dacay-rates}(a) that in the OPB\ regime $\Gamma _{1+,1-}$ is
suppressed while $\Gamma _{1+,0}$ and $\Gamma _{1-,0}$ are always finite.
The transition to the C-TPB occurs around $g\simeq 0.253\omega _{0}$
exceeding which $\Gamma _{1+,1-}$ arises from the vanishing ones. In
contrast, the U-TPB is located in the vanishing-$\Gamma _{1+,1-}$ regime,
with this primary cascade channel closed.

Now in the presence of the Stark coupling, as shown in Fig.~\ref%
{Fig-dacay-rates}(b), the enhanced U-TPB (pale-rose region) also has the
closed cascade channel, as the broadened coupling window is entirely inside
the vanishing-$\Gamma _{1+,1-}$ regime.

Correspondingly the population on the state $|\psi _{1-}\rangle $ in the
U-TPB\ would be vanishing, due to the closed cascade decay channel,
differently from the finite population in the C-TPB with the opened cascade
channel. Indeed, as shown in Fig.\ref{Fig-population}, the population on
$|\psi _{1-}\rangle $ (dashed line) is vanishing in the U-TPB phase both in
the absence and presence of the Stark coupling.

In Fig.~\ref{Fig-population} we also plot the energy difference $\Delta _{12}$
(dotted line). Although $\Delta _{12}$ differs much in the absence and
presence of the Stark coupling, the vanishing $|\psi _{1-}\rangle $
population is unaffected in the U-TPB, due to that no population transits from the driven state $|\psi _{1+}\rangle $
to this primary decay state $|\psi _{1-}\rangle $ in the closed decay channel.
Nevertheless, the difference of $\Delta _{12}$ will influence the decay rate
[see Eq.(\ref{Eq-Decay-Rate})] in the C-TPB where the transition matrix
element $C_{jk}^{1-,1+}$ becomes finite, thus providing a channel parameter
for the manipulation of the C-TPB.

\subsection{Weak anharmonicity and non-vanishing population above the
cascade levels}

The U-TPB occurs in the situation of weak anharmonicity $\Delta $ as defined
in (\ref{Eq-anharmon}). In such a situation, as the energy difference
$(E_{2,+}-E_{1,+})\ $is similar to $(E_{1,+}-E_{0})$, in the driving from
$|\psi _{0}\rangle $ to $|\psi _{1+}\rangle $, there is also some population
on $|\psi _{2+}\rangle $ indirectly driven from $|\psi _{1+}\rangle $ due to
the resonance effect. This additional population opens new channel for the
two-photon process. On the other hand, this additional population is small and not able to support a more-photon process,
as population diminishes in each channel transition and higher-order photon process with further channel transition would appear to be vanishing, which keeps the three-photon process still blockaded.  As a consequence, the two-photon correlation function $g^{(2)}(0)$ increases but the three-photon one $g^{(3)}(0)$ remains suppressed,
thus giving rise to the U-TPB.

We can see in Fig.~\ref{Fig-anharmonicity} that the anharmonicity $\Delta $
(blue dotted line, blue right axis) is indeed small in the U-TPB regime both
in the absence (panel (a)) and presence (panel (b)) of the Stark coupling,
while it is large in the C-TPB phase. In Fig.~\ref{Fig-anharmonicity} we can
also observe that the population on $|\psi _{2+}\rangle $ (solid line) is indeed
non-vanishing but not small relatively to those of the main cascade levels in Fig.~\ref{Fig-population}.
We also find that with the Stark coupling population on
$|\psi _{2+}\rangle $ maintains in the similar values in a wide coupling
regime, while without the Stark coupling this population grows quickly in coupling decreasing or diminish soon in the coupling increasing.
This appropriate-population maintaining may be the favorable situation for the broadened U-TPB regime
in the Stark coupling.

\begin{figure}[t!]
	\centering
    \includegraphics[width=0.88\linewidth]{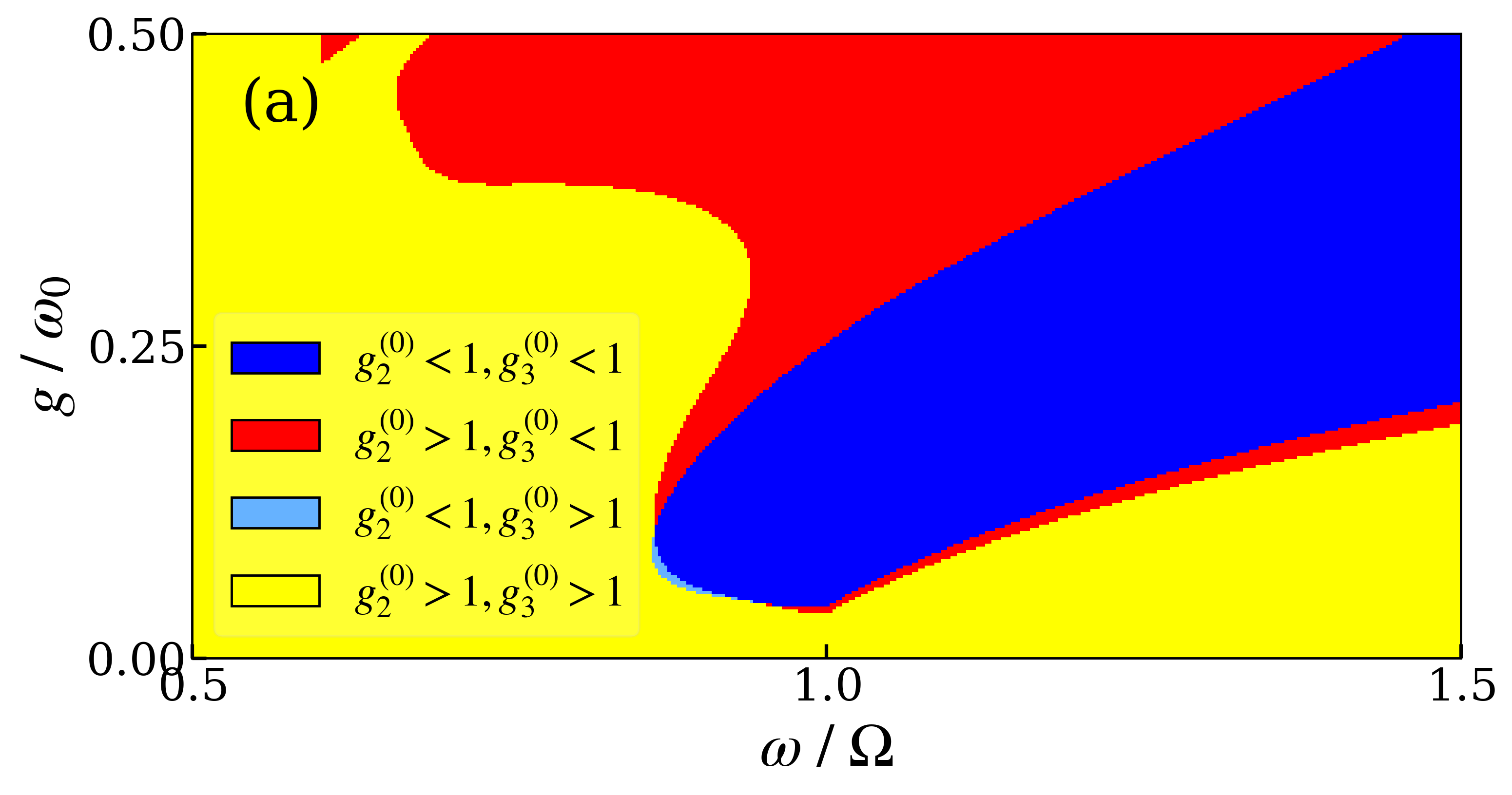}
	\includegraphics[width=0.88\linewidth]{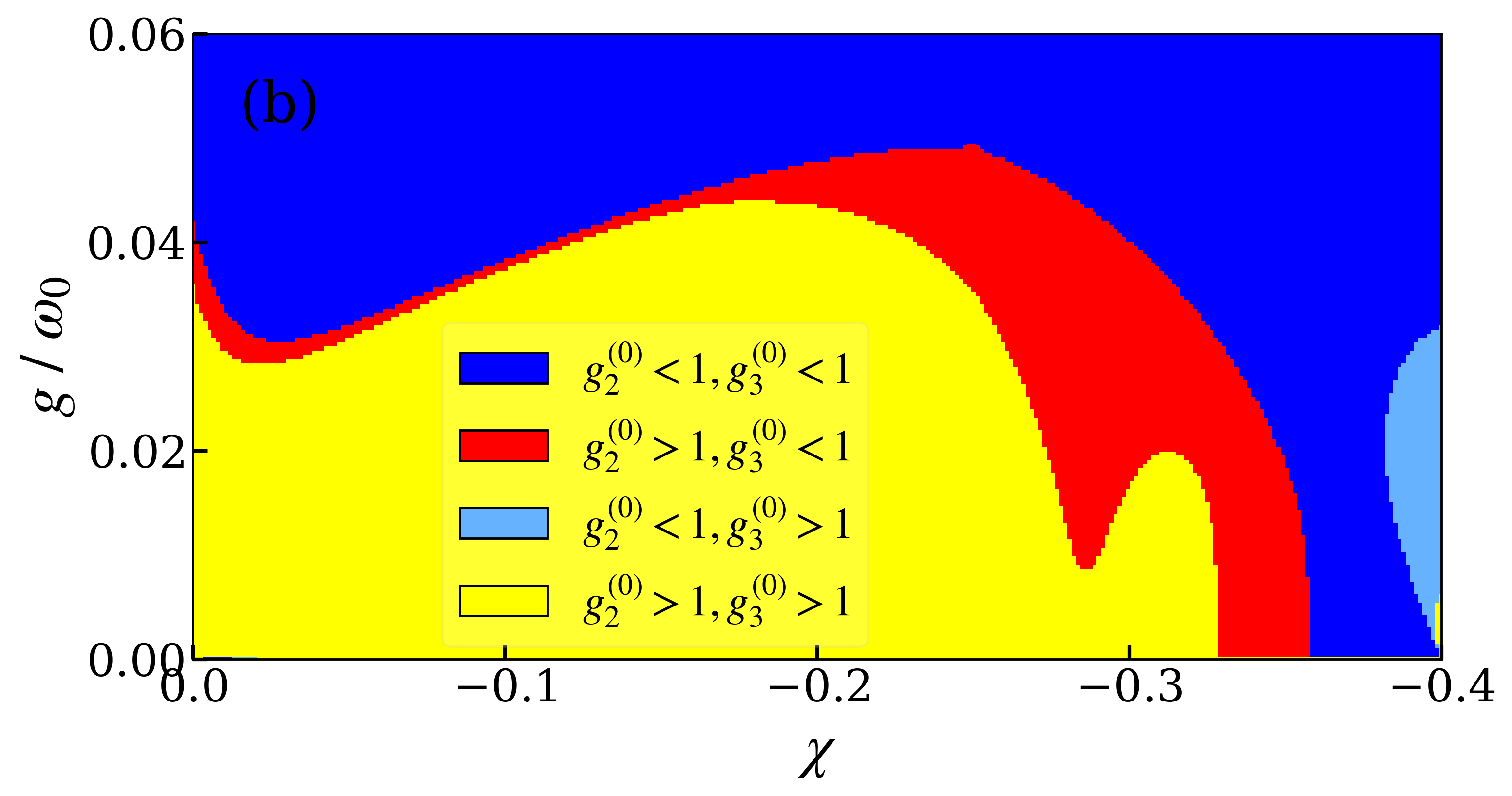}
	\includegraphics[width=0.88\linewidth]{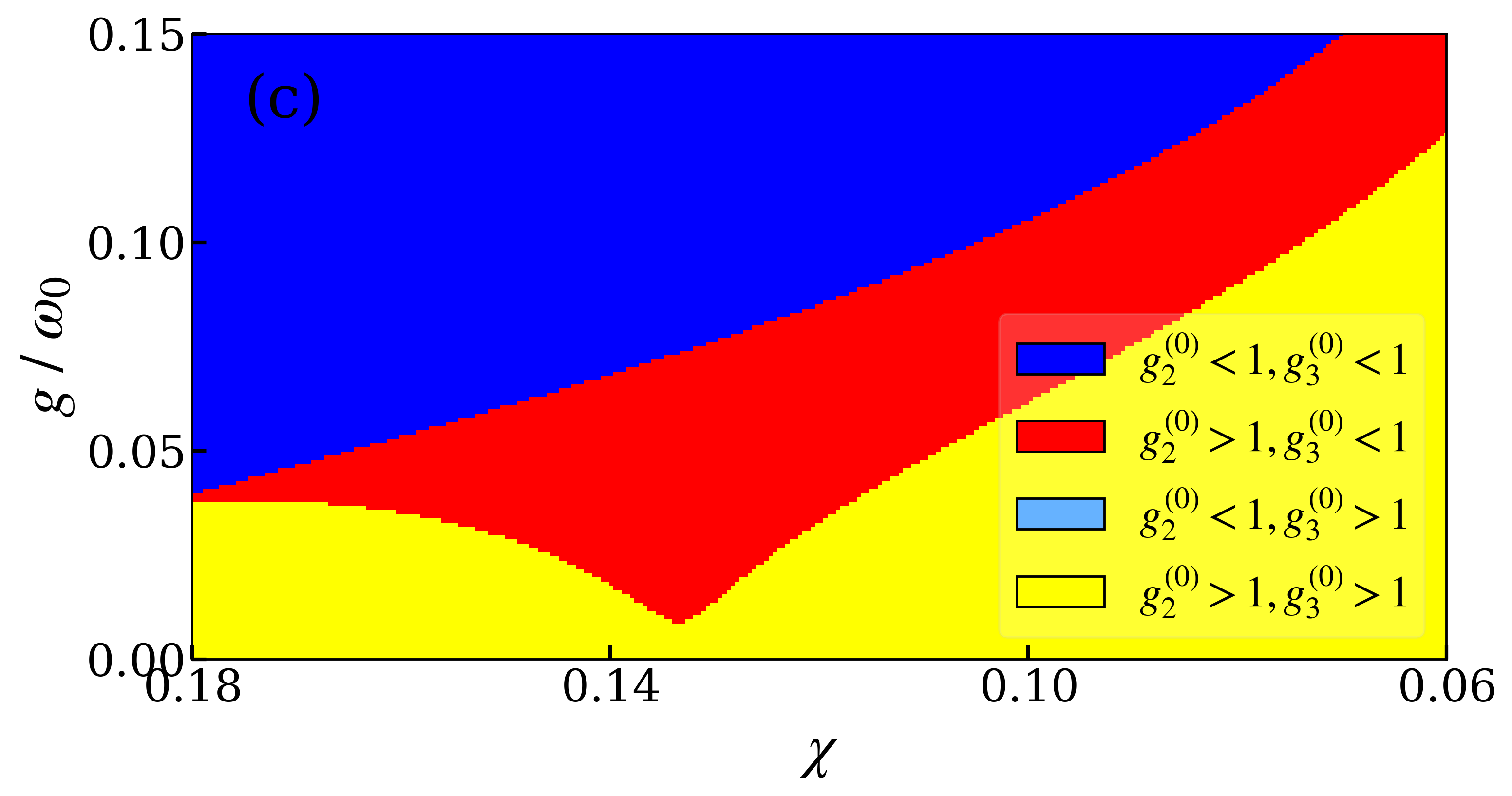}
	\caption{Phase diagrams for enhanced U-TPB by Stark nonlinear coupling $\chi$:
(a) $\chi =0$, without the Stark coupling the U-TPB (red slice) is not much enhanced by the variation of frequency ratio $\omega/\Omega$. (b) $\omega=1.0\omega_0$, the U-TPB (red) is dramatically enhanced by the Stark coupling $\chi$. (c) $\omega=1.5\omega_0$, combination of the Stark coupling and detuning enables enhancement covering the entire strong coupling regime. Here, different colors correspond to non-blockading (yellow) with $g^{(2)}(0) >1$ and $g^{(3)}(0)>1$, OPB (blue) with $g^{(2)}(0) <1$ and $g^{(3)}(0)<1$, and TPB phase (red) with $g^{(2)}(0) >1$ and $g^{(3)}(0)<1$.
}
\label{Fig-diagram-enhance}
\end{figure}

\section{Phase diagrams of enhanced U-TPB}

\label{Section-diagram-enchance}

To have an overview of the enhanced U-TPB, we compare the phase diagrams of $%
g^{(2)}(0)$ and $g^{(3)}(0)$ in the presence and absence of the Stark
coupling. In the phase diagrams, non-blockading phase ($g^{(2)}(0)>1$ and $%
g^{(3)}(0)>1$), OPB ($g^{(2)}(0)<1$ and $g^{(3)}(0)<1$) and TPB ($%
g^{(2)}(0)>1$ and $g^{(3)}(0)<1$) are marked in yellow, blue and red
respectively. 

\subsection{Manipulation by detuning: Effective for C-TPB but inefficient for
U-TPB}

We first check the case without the Stark coupling. The examples in
the last Section are illustrated at resonance $\omega =\Omega $. Without the
Stark coupling we still can tune the different phases via detuning by going
away from the resonance situation. As shown in Fig.~\ref{Fig-diagram-enhance}(a),
when we vary the frequency ratio $\omega /\Omega $, the boundary of
C-TPB phase (broad red region above $g\sim 0.2\omega _{0}$) can be tuned
effectively, with the critical coupling nearly linearly increasing for $\omega >\Omega $ and
non-monotonously changing for $\omega <\Omega $.
However, the U-TPB phase (red slice below $g\sim 0.2\omega _{0}$) remains
narrow in the variation of the frequency ratio, except for some slight broadening and the
position moving. These results indicate that without the Stark coupling
it is not efficient to enhance the U-TPB solely by detuning.

\subsection{Manipulation by Stark coupling: Dramatic enhancement for U-TPB}

In Fig.~\ref{Fig-diagram-enhance}(b), we show the phase diagram in variation
of the Stark coupling $\chi $ at resonance $\omega =\Omega $ without resorting to detuning. Here the red
region is entirely the U-TPB phase. We see that at $\chi =0$ the U-TPB phase
is narrow, as expected. However, when we tune the Stark coupling to the
regime $-0.2\gtrsim g/\omega _{0}\gtrsim -0.35$, the U-TPB phase is
dramatically broadened. This result demonstrates that the Stark coupling can
be an effective manipulation for the enhancement of the U-TPB.

\subsection{Combination of Stark coupling and detuning: Enhancement
covering the entire strong-coupling regime}

We an combine the Stark coupling\ and detuning to have more flexibility for
the U-TPB enhancement. Figure \ref{Fig-diagram-enhance}(c) illustrates the
phase diagram at $\omega =1.5\omega _0 $. In this situation, varying the Stark
coupling we have a broad U-TPB which is moving and capable of covering the
entire strong-coupling regime ($0.01\lesssim g/\omega _{0}\lesssim 0.1$).

\begin{figure}[t!]
	\centering
    \includegraphics[width=0.88\linewidth]{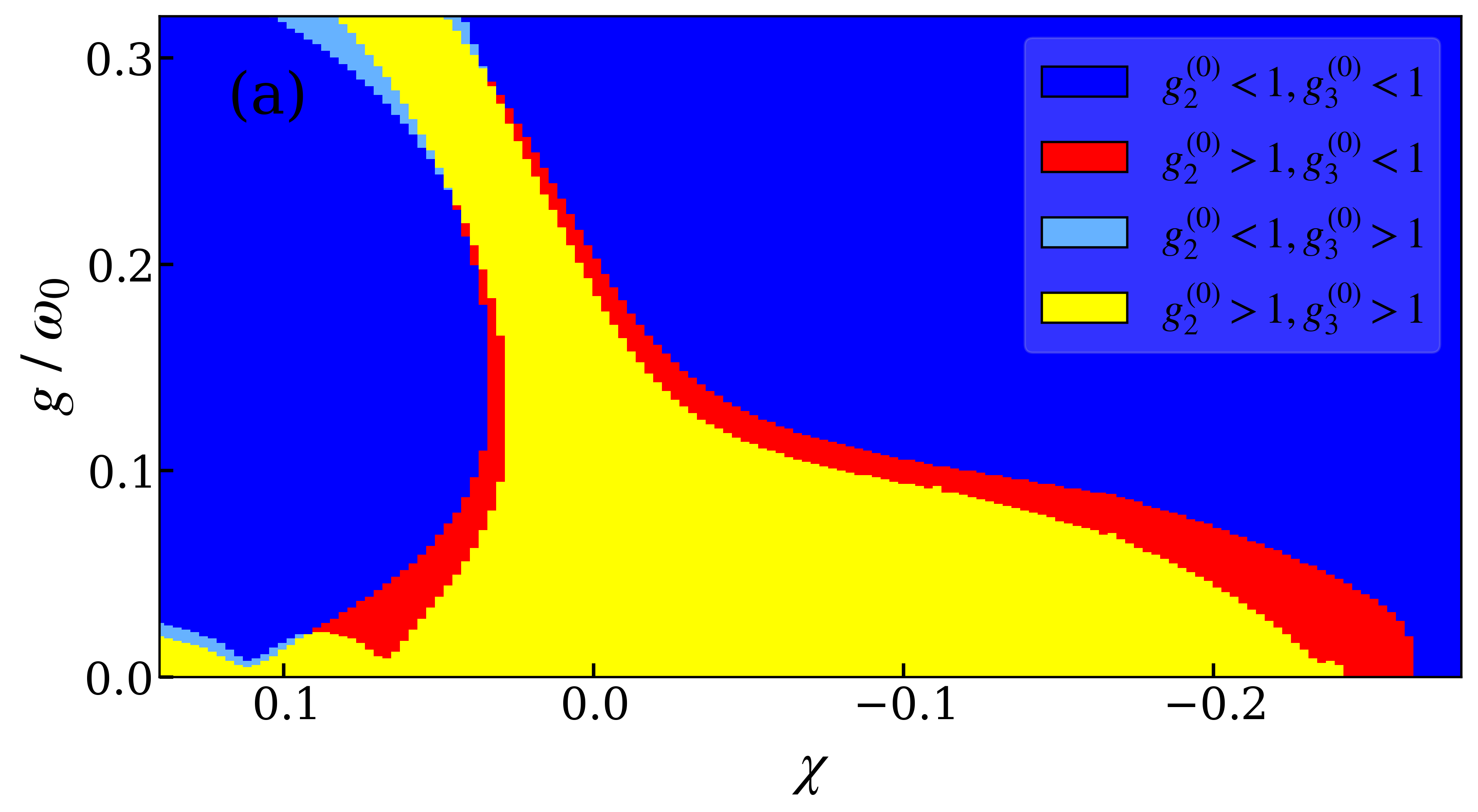}
	\includegraphics[width=0.88\linewidth]{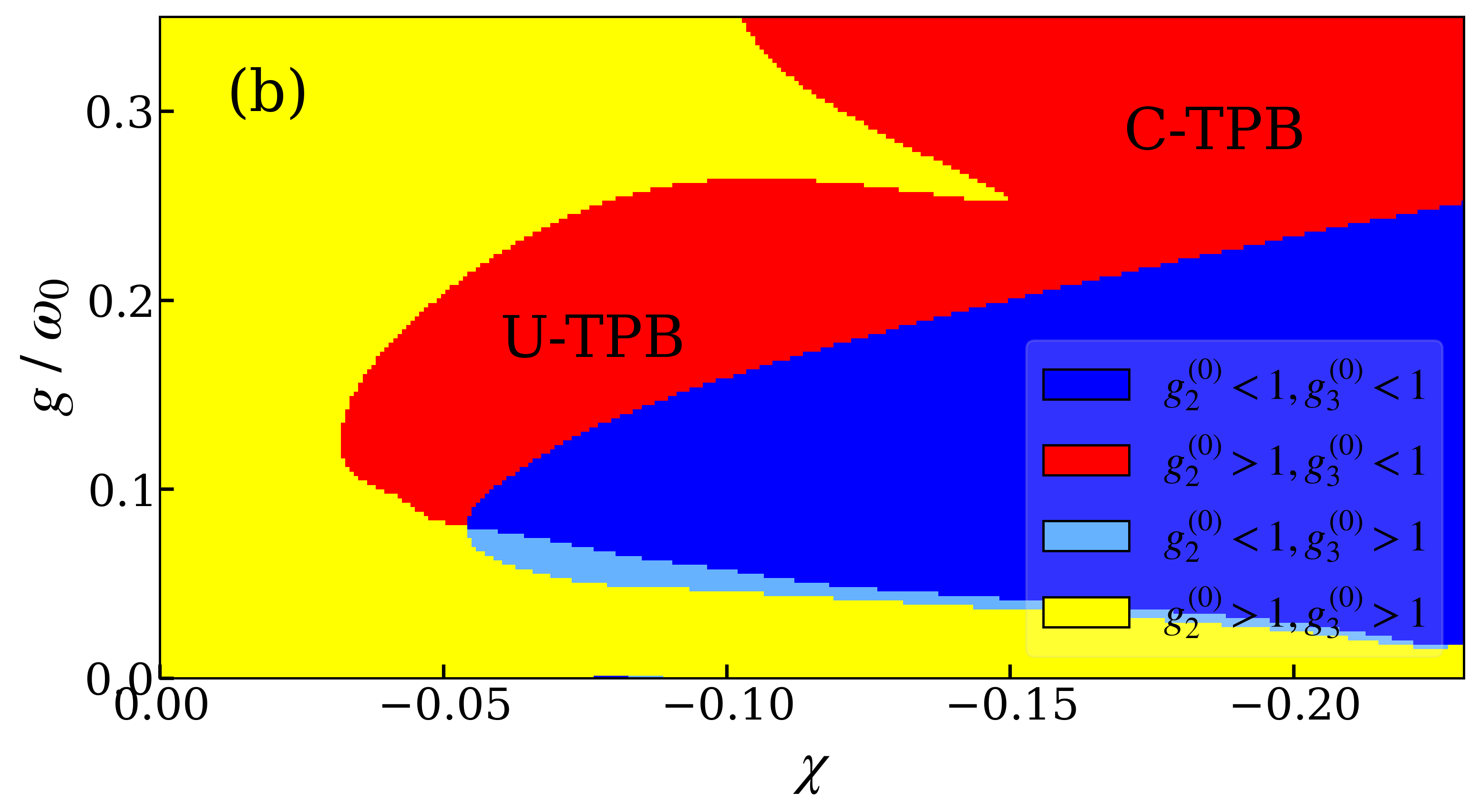}
	\caption{(a) Two U-TPB enhancement branches (red belts) at $\omega=1.2\omega_0$.  (b) Broad U-TPB in ultrastrong coupling and crosover to the C-TPB at $\omega=0.9\omega_0$.}
\label{Fig-diagram-crossover}
\end{figure}

\begin{figure}[t!]
	\centering
    \includegraphics[width=0.88\linewidth]{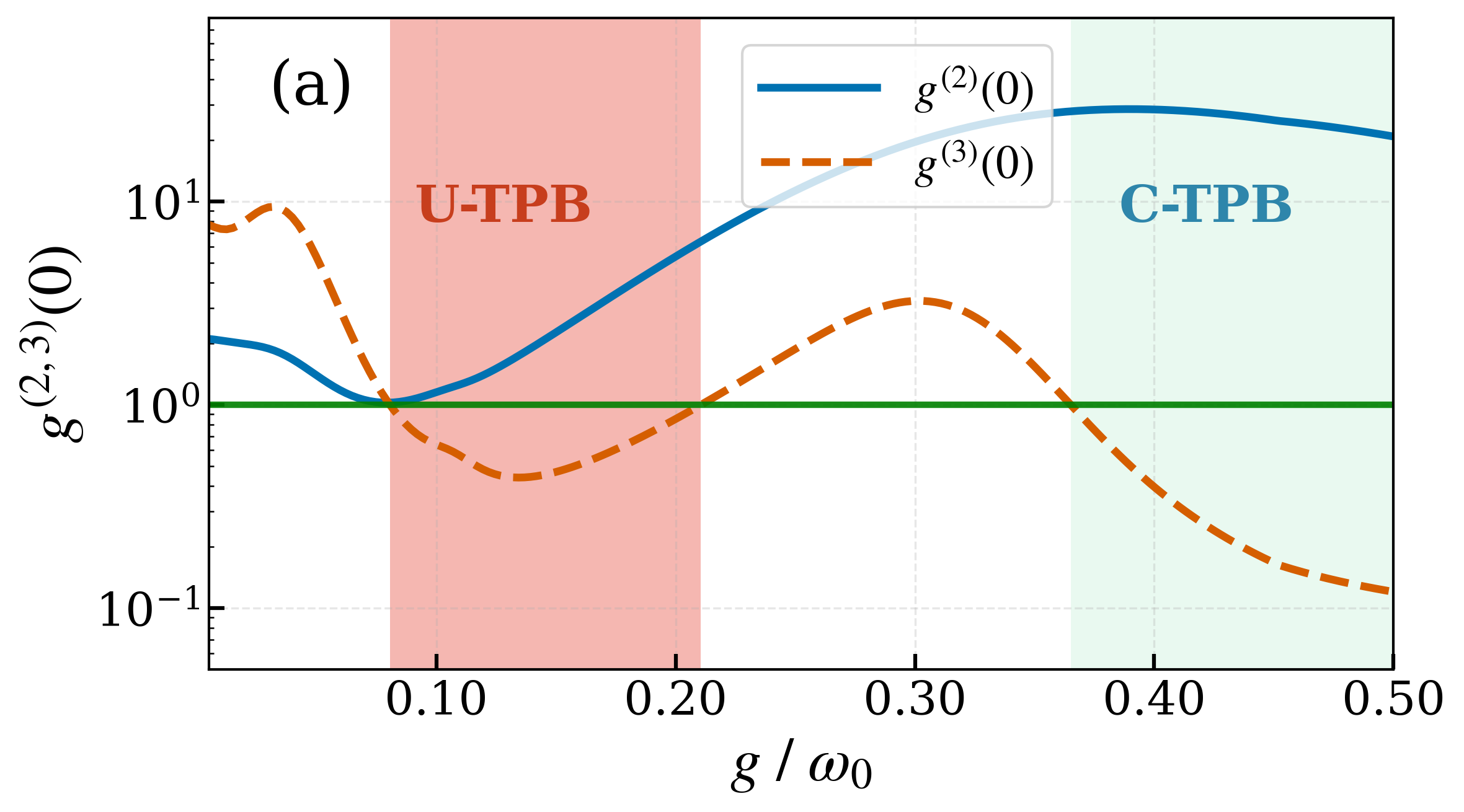}
	\includegraphics[width=0.88\linewidth]{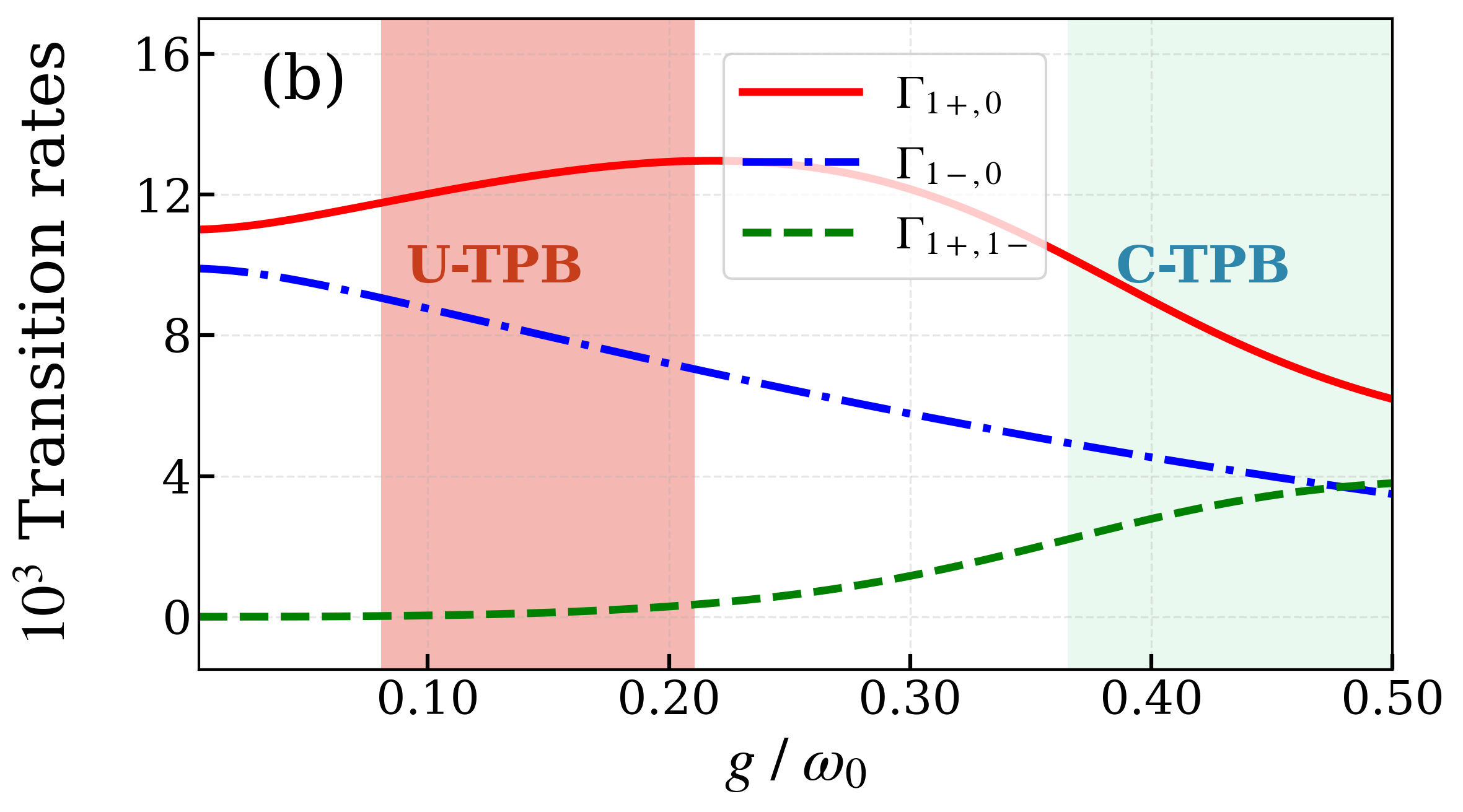}
	\caption{Photon correlation functions (a) and decay transition rates (b) for the U-TPB in ultrastrong coupling, at $\omega=0.9\omega_0$ and $\chi =-0.053$.}
\label{Fig-G2G3-ultrastrong}
\end{figure}

\section{Broad U-TPB in ultrastrong coupling and crossover to C-TPB}
\label{Section-diagram-crossover}

\subsection{Above the resonance ($\protect\omega >\Omega $): Two U-TPB
enhancement branches and a wide U-TPB in ultrastrong coupling}

The enhanced U-TPB phases in Figs.\ref{Fig-diagram-enhance}(b) and \ref{Fig-diagram-enhance}(c)
actually belong to two enhancement branches in the
negative-$\chi $ and positive-$\chi $ regimes. In Fig.\ref%
{Fig-diagram-crossover}(a) we show a more overall view of the two
enhancement branches at $\omega =1.2\omega_0 $. We see that both U-TPB
branches (red belts) are covering the entire strong-coupling regime\ and
have the tendency of extending to the ultrastrong coupling regime ($g/\omega
_{0}\gtrsim 0.1$). For a higher frequency the two
branches will span over a larger range of $\chi $.

Particularly in the positive-$\chi $ branch, in the coupling regime shifting
from the strong coupling to the ultrastrong coupling, there is returning
point in the variation of $\chi $. For $\omega =1.2\Omega $ in Fig.\ref{Fig-diagram-crossover}(a)
this returning point appears around $\chi \sim
0.033$. In the Stark-coupling window $\chi \in (0.027,0.034)$, an U-TPB opens
wich can be vertically wide along the $g$ direction. For an example, at $\chi =0.033$ the
U-TPB spans over the coupling range $g\in (0.080,0.0184)$ which is also much
wider relatively to the case in the absence of the Stark coupling as in Fig.~\ref{Fig-g2g3-enhanced}. Also special is that the U-TPB is entering the
ultrastrong coupling regime. Actually, this $\chi$-returning point gives a further
widening of the U-TPB as the U-TPB phase evolves and moves to the ultrastrong coupling regime from the strong-coupling regime.

\subsection{Below the resonance ($\omega <\Omega $): A broad U-TPB
in ultrastrong coupling and crossover to C-TPB}

The above two-branch scenario occurs above the resonance ($\omega >\Omega $), while below the resonance ($\omega <\Omega $) we find another broad U-TPB
region in the ultrastrong coupling regime, with a crossover to C-TPB phase.
Figure \ref{Fig-diagram-crossover}(b) displays such a phase diagram in the
negative--$\chi $ regime at $\omega =0.9\Omega $. The broadest red area in
the regime $g\gtrsim 0.23$ is the C-TPB phase. Here we also see the
tunability of C-TPB phase by the Stark coupling, with a direct reason from
the tunable level gaps $\Delta _{jk}$ [e.g. Eq.~(\ref{Eq-gap12})] which
directly control the decay transition rates $\Gamma _{jk}$ in Eq.~(\ref{Eq-Decay-Rate}),
though our focus is on the U-TPB phase. We find an U-TPB
phase in the area $-0.03\gtrsim \chi \gtrsim -0.12$ and $0.1\lesssim
g/\omega _{0}\lesssim 0.25$. This U-TPB phase opens around $\chi \thicksim
-0.03$ at $g/\omega _{0}\thicksim 0.135$ with suppressed decay rate $\Gamma
_{1+,1-}$ and soon expands to a wide range $0.08\lesssim g/\omega
_{0}\lesssim 0.2$. As an illustration, we plot $g^{(2)}(0)$ and $g^{(3)}(0)$
in Fig.~\ref{Fig-G2G3-ultrastrong}(a) and the decay transition rates in Fig.~\ref{Fig-G2G3-ultrastrong}(b) at $\chi =0.053$. Here, the broad U-TPB phase
is still well separated from the C-TPB, with a suppressed primary cascade
decay channel $\Gamma _{1-,1+}$ as shown by the dashed line in Fig.~\ref{Fig-G2G3-ultrastrong}(b). This U-TPB window moves towards large coupling as
the $\chi $ is tuned to be more negatively stronger. At the same time, the
U-TPB and C-TPB phases get closer and finally meet around $\chi \thicksim
-0.15$. In this process, $\Gamma _{1-,1+}$ gradually arises and finally
reaches the finite values as a C-TPB state. Thus, the enhanced U-TPB in the
ultra-strong coupling regime is connected to the C-TPB phase via a crossover.

Finally it should be mentioned that there also emerges a phase with $g^{(2)}(0)<1$ and $g^{(3)}(0)>1$ (light-blue) in
Fig.~\ref{Fig-diagram-enhance}(b) and Fig.~\ref{Fig-diagram-crossover}. Such a peculiar
case can occur in several different situations. As it is beyond our focus on
the enhancement of the U-TPB and the mechanisms are more involved, we leave
for a special discussion elsewhere.

\section{Conclusions}
\label{Section-Conclusion}

In this work, we have proposed to enhance the U-TPB by the Stark nonlinear
coupling. The U-TPB found in the linear coupling is limited in a narrow coupling
window and the blockading strength is weak. We demonstrated that adding the
Stark nonlinear coupling is capable of not only broadening the coupling
window of U-TPB but also deepening the blockading degree.

We first carried out an analytical analysis in the leading order of small
coupling in the presence of the Stark coupling, which shows that the Stark
coupling can tune the weak anharmonicity while keeping the symmetries of
parity and excitation number. While the weak anharmonicity
plays a key role in formation of the U-TPBThe, the symmetry preserving keeps the vanishing
zeroth order of decay transition element which lays the characteristic condition
of closed cascade decay channel for the U-TPB.

After a brief introduction of the input-output theory for photon blockading,
we illustrated a typical enhancement example in the presence of the Stark coupling.
Indeed, by checking the photon correlation functions $g^{(2)}(0)$ and $g^{(3)}(0)$,
we found the coupling window of U-TPB is dramatically broadened
in comparison to the case without the Stark coupling. Moreover, the
blockading degree of U-TPB is much deepened, actually with the TPB
blockading quantity $g^{(3)}(0)$ reduced by two orders. We also verified the
vanishing transition rate $\Gamma _{1+,1-}$ and vanishing population of
state $|\psi _{1+}\rangle $, which demonstrates the unconventional closure of
the primary cascade channel as the character of the U-TPB distinguished from the C-TPB
conventionally with open cascade channel. We showed that the weak
anharmonicity is also present in the presence of the Stark coupling and the
population on $|\psi _{2+}\rangle $ is non-vanishing. In fact, the weak
anharmonicity leads to a resonance effect in driving and induces a
non-vanishing population of $|\psi _{2+}\rangle $ which is above the closed
cascade channel. The population of $|\psi _{2+}\rangle $ opens a new channel
to raise $g^{(2)}(0)$ that gives rise to the U-TPB, in contrast to the
opening of cascade channel in the formation of the C-TPB. We found the
population of $|\psi _{2+}\rangle $ remains around a similar appropriate value, non-vanishing for raising $g^{(2)}(0)$
but small for keeping $g^{(3)}(0)$ suppressed, in a wide
coupling range, which may be favorable for the window broadening of the U-TPB.

We further compared phase diagrams in the absence and presence of the Stark
coupling. We showed that frequency detuning can effectively tune the phase
boundary of C-TPB but is not efficient for the enhancement of the U-TPB except
for position moving. In contrast, the Stark coupling can much broaden the
U-TPB phase even without the frequency detuning. In spite of their different trends in tuning, the
combination of the Stark coupling and the frequency detuning can help to
optimize the U-TPB to cover the entire strong-coupling regime.

Finally we provided a more overall view of the phase diagram which
summarizes the two enhancement branches of U-TPB above the resonance. Here, a further widening of U-PTB occurs around the $\chi$-returning point.
While the other enhanced U-PTBs mainly
lie in the strong coupling regime, below the resonance we also revealed a broad phase
of U-TPB in the ultrastrong coupling regime, with a connection to the C-TPB by a crossover.

Although our proposal has been focusing on the enhancement of the U-TPB~\cite{Han2026WeakTPB},
the Stark coupling can also be an effective control
parameter on the OPB and C-TPB as indicated in Figs.~\ref{Fig-diagram-enhance}
and \ref{Fig-diagram-crossover}. Note that the
Stark coupling~\cite{Eckle-2017JPA,*Eckle-2017JPA-b,Stark-Cong2020,Ying-Stark-top,*Ying-Stark-top-Cover}
can be realized and tailored~\cite{Stark-Grimsmo2013,Stark-Grimsmo2014,Stark-Cong2020,Zhai2025TwoPhotonStark},
not only with large flexibility in the coupling strength but also variable
in the coupling sign~\cite{Stark-Grimsmo2013,Stark-Grimsmo2014}. It is also
worth mentioning that the relevant Stark coupling values in the present work
are all within the stable range $\chi \in \lbrack -1,1]$ \cite%
{Ying-Stark-top,*Ying-Stark-top-Cover}. Our proposal may pave a practical
way not only for enhancement of the U-TPB but also for manipulation of
different types of photon blockades~\cite{Boite2016-Photon-Blockade,Ridolfo2012-Photon-Blockade,LiaoJQ2020BlockadeJC,Ma2026PRL-Strong-2PB,Felicetti2026PRXQuantumBlockade,
Carmichael1985,Birnbaum2005,Shamailov2010,Liew2010,Hamsen2017,Garziano2017,Flayac2017},
which may be useful for development of the potential quantum technologies
in manipulation of light on the level of individual quanta.

\section*{Acknowledgments}

This work was supported by the National Natural Science Foundation of China
(Grants No. 12474358, No. 11974151, and No. 12247101).

\bibliography{Refs-2026-6-2Photon-Blockade}

\end{document}